\documentclass[
 ,secnumarabic%
,showpacs ,amssymb,nobibnotes, aps]{revtex4}
\usepackage{amsmath}
\usepackage{graphicx}
\input{epsf}

\usepackage{amsmath,amssymb}
\usepackage{bm}

\newcommand{\dalm}{\kern1pt\vbox{\hrule height 0.9pt\hbox{\vrule width 0.9pt\hskip 2.5pt\vbox{\vskip 5.5pt}\hskip 3pt\vrule width 0.3pt}\hrule height 0.3pt}\kern1pt}

\begin{document}



\title{Probing Strong-Field Scalar-Tensor Gravity with Gravitational Wave Asteroseismology}
\author{Hajime Sotani}\email{sotani@gravity.phys.waseda.ac.jp}
\affiliation{Research Institute for Science and Engineering, Waseda University,\\
Okubo 3-4-1, Shinjuku, Tokyo 169-8555, Japan}

\author{Kostas D. Kokkotas}\email{kokkotas@auth.gr}
\affiliation{Department of Physics, Aristotle University of
Thessaloniki, Thessaloniki 54124, Greece
\\ Center for Gravitational Wave Physics, 104 Davey Laboratory, University Park, PA 16802, USA.
}

\date{\today}

\begin{abstract}
We present an alternative way of tracing the existence of a scalar field based on the analysis of the gravitational wave spectrum of a vibrating neutron star. Scalar-tensor theories in strong-field gravity can potentially introduce much greater differences in the parameters of a neutron star than the uncertainties introduced by the various equations of state. The detection of gravitational waves from neutron stars can set constraints on the existence and the strength of scalar fields.
We show that the oscillation spectrum is dramatically affected by the presence of a scalar field, and can provide unique confirmation of its existence.
\end{abstract}

\maketitle
\section{Introduction}
\label{sec:Intro}

A natural alternative/generalisation to general relativity is the
scalar-tensor theory in which gravity is mediated by long-range
scalar fields in addition to the usual tensor field present in
Einstein's theory
\cite{Fierz1956,Jordan1959,Brans1961,Damour1992,Will1993,Will2001}.
Scalar-tensor theories of gravity can be obtained from the
low-energy limit of string theory or/and other gauge theories. The
existence of scalar fields is crucial in explaining the
accelerated expansion phases of the universe e.g. inflation and
quintessence. They are viable theories of gravity for a specific
range of the function that couples the scalar field to gravity.
 Still it is not clear how the scalar fields couple to gravity
\cite{Damour1994a,Damour1994b}, a basic assumption is that the
scalar and gravitational fields $\varphi$ and $g_{\mu\nu}$ are
coupled to matter via an ``effective metric'' ${\tilde
g}_{\mu\nu}= A^2(\varphi)g_{\mu\nu}$. The existence of the scalar
field has not yet been verified while a number of experiments in
the weak field limit of general relativity disproved or set severe
limits in the existence and the strength of the scalar
fields \cite{Will2001,Esposito2004}.

The Fierz-Jordan-Brans-Dicke \cite{Fierz1956,Jordan1959,Brans1961}
theory assumes that the ``coupling function'' has the form
$A(\varphi)=\alpha_0 \varphi$ i.e. it is characterized by a unique
free parameter $\alpha_0^2=(2\omega_{\rm BD}+3)^{-1}$ and all its
predictions differ from those of general relativity by quantities
of order $\alpha_0^2$ \cite{Damour1993}. Solar system experiments
set strict limits in the value of the Brans-Dicke parameter
$\omega_{\rm BD}$ i.e. $\omega_{\rm BD}\agt40000$ which suggests a
very small $\alpha_0^2<10^{-5}$ see \cite{Esposito2004,Bertotti2003,Wiaux-1,Wiaux-2}.
Damour and Esposito-Farese \cite{Damour1993,Damour1996} showed
that the predictions of scalar-tensor theories in the strong field
might be drastically different from those of general relativity.
By studying neutron star models in a simplified version of scalar
tensor theory where $A(\varphi)=\alpha_0\varphi+\beta \varphi^2/2$
they found that for certain values of the coupling parameter
$\beta$ the stellar models develop some strong field effects which
induce significant deviations from general relativity. Their claim
was based on the difference that one can observe in the properties
of neutron stars by the introduction of a scalar field (even if
the coupling constant $\alpha_0$ is very small). Damour and
Esposito-Farese described the sudden deviation from general
relativity for specific values of the coupling constants as
``spontaneous scalarization". Harada \cite{Harada1998} studied in
more detail the stability of non-rotating neutron stars in the
framework of the scalar-tensor theory and he reported that
``spontaneous scalarization" is possible for $\beta\alt-4.35$.
In other words the ``spontaneous scalarization'' of Damour and
Esposito-Farese suggests that weak-field experiments cannot
constrain the effect of the scalar fields for the strong-field
regime and prompts for alternative measurements. Such measurements
can be based, for example, on accurate estimation of the orbital
decay of binary systems \cite{Damour1996,Damour1998}, in the
accurate monitoring of gravitational waveforms from neutron stars
spiralling into massive black holes \cite{Scharre2002} or by
direct observation of monopolar gravitational radiation during the
collapse of compact objects. Collapse simulations have shown that
indeed a scalar field can be observed by LIGO/EGO for the specific
range of values of $\beta$ if the event takes place in our Galaxy
\cite{Shibata1994,Scheel1995,Harada1997,Saijo1997,Novak1998,Novak2000},
while the space gravitational wave detector LISA can also provide
constraints in the existence of scalar field \cite{Scharre2002,Will2004}.

Recently, DeDeo and Psaltis \cite{DeDeo2003} suggested that the
effects of the scalar fields might be apparent in the observed
redshifted lines of the X-rays and $\gamma$-rays observed by
Chandra and XMM-Newton. Testing strong gravity via electromagnetic
observations is a really novel idea, which have been recently
extended, by the same authors \cite{DeDeo2004}, in suggesting
tests via the quasi-periodic oscillations (QPOs).

In this paper we examine whether gravitational wave observations of
the oscillation spectra of neutron stars can provide an
alternative way of testing scalar-tensor theories. The aforementioned
results \cite{Damour1993,Harada1998,DeDeo2003} show that for a
specific range of values of the coupling constant $\beta$, which
are not constrained by the current experimental limits in the
weak-field regime, neutron stars can have significantly larger
masses and radii. This immediately suggests that the natural
oscillation frequencies of the neutron stars will be altered
accordingly and a possible detection of gravitational waves from
such oscillations will not only probe the existence of the scalar
field but might provide a way of estimating its strength.

The estimation of the stellar parameters (mass, radius and
equation of state) via their oscillation properties is not a new
idea. Helioseismology and asteroseismology are established fields
in Astronomy and there is already a wealth of information about
the interior of our sun and the stars via this approach. In the
late `90s it was  suggested \cite{Andersson1996,Andersson1998}
that the oscillation spectra of neutron stars can reveal in a
unique way their properties. The radius, the mass and the equation
of state can be easily deduced by an analysis of the oscillation
spectrum of the $f$, $p$ and $w$-modes \cite{Kokkotas2001}.
Moreover, the rotation period of a neutron star can be revealed by
the $r$-mode oscillations since their frequency is proportional to
the rotation rate ($\omega_{\rm r-mode} \sim 4\Omega/3$). Features
such as superfluidity \cite{Andersson2001a} or magnetic fields
\cite{Kokkotas2004} can reveal their presence via detailed
analysis of the gravitational and electromagnetic spectra.
Recently, it has been suggested that compact stars with exotic
equations of state, such as strange stars, have a spectrum which
carries in a unique way their signature
\cite{Sotani2003,Sotani2004}.

The oscillations of a neutron star in the  scalar-tensor theory
will produce not only gravitational but also scalar
waves \cite{Will1993}. Detectable scalar waves will be a unique
probe for the theory, but this might not be the case if the
radiated energy is small. Still, we show that there is no need for
direct observation of scalar waves, the presence of the scalar
field will be apparent in the gravitational wave spectrum, since
the spectra will be shifted according to the strength of the
scalar field.

The paper is organized as follows. In the next section we present
the basic equations for the construction of the unperturbed
spherically symmetric stellar models. We also show the effect of
the scalar field on the stellar structure. In Section
\ref{sec:III} we derive  the perturbation equations which will be
used  for the numerical estimation of the oscillation frequencies.
Finally, in Section \ref{sec:conclusions} we discuss the results
in connection to gravitational wave asteroseismology. Note that
the present study will be based on two equations of state, which
have been used  earlier \cite{Damour1993,Harada1998}. Our aim is
to demonstrate the effect of the scalar field in the neutron star
oscillation spectra, while more detailed analysis for a wide range of
EOS is underway.

\section{Stellar models in scalar-tensor theories of gravity}
\label{sec:II}

In this section we will study neutron star models in scalar-tensor
theory of gravity with one scalar field. This is a natural
extensions of Einstein's theory, in which gravity is mediated
not only by a second rank tensor (the metric tensor $g_{\mu\nu}$)
but also by a massless long-range scalar field $\varphi$. The
action is given by \cite{Damour1992}
\begin{equation}
 S = \frac{1}{16\pi G_*}\int\sqrt{-g_*}
   \left(R_*-2g_*^{\mu\nu}\varphi_{,\mu}\varphi_{,\nu}\right)d^4x
   + S_m\left[\Psi_m,A^2(\varphi)g_{*\mu\nu}\right],
\label{eq:action}
\end{equation}
where all quantities with asterisks are related to the ``Einstein
metric'' $g_{*\mu\nu}$, then $R_*$ is the curvature scalar for this
metric and $G_*$ is the bare gravitational coupling constant.
$\Psi_m$ represents collectively all matter fields, and $S_m$
denotes the  action of the matter represented by $\Psi_m$, which
is coupled to the ``Jordan-Fierz metric tensor''
$\tilde{g}_{\mu\nu}$. The field equations formulated better in the
``Einstein metric'' but all non-gravitational experiments measure
the ``Jordan-Fierz'' or ``physical metric''. The ``Jordan-Fierz metric''
is related to the ``Einstein metric'' via the conformal
transformation,
\begin{equation}
 \tilde{g}_{\mu\nu} = A^2(\varphi)g_{*\mu\nu}.  \label{metric-relation}
\end{equation}
Hereafter, we relate all tilded quantities with  ``physical
frame'' and those with asterisk with the  ``Einstein frame''.
From the variation of the action $S$ we get the field equations in the Einstein frame
\begin{align}
 G_{*\mu\nu} &= 8\pi G_*T_{*\mu\nu}
              + 2\left(\varphi_{,\mu}\varphi_{,\nu}-\frac{1}{2}g_{*\mu\nu}g_*^{\alpha\beta}
                \varphi_{,\alpha}\varphi_{,\beta}\right), \label{basic1} \\
 \dalm_*\varphi &= -4\pi G_*\alpha(\varphi)T_*, \label{basic2}
\end{align}
where $T_*^{\mu\nu}$ is the energy-momentum tensor in the Einstein frame
which is related to the physical energy-momentum tensor $\tilde{T}_{\mu\nu}$ as follows,
\begin{equation}
 T_*^{\mu\nu} \equiv \frac{2}{\sqrt{-g_*}}\frac{\delta S_m}{\delta g_{*\mu\nu}}= A^6(\varphi)\tilde{T}^{\mu\nu}.
\end{equation}
The scalar quantities $T_{*}$ and $\alpha (\varphi)$ are defined as
\begin{align}
 T_* &\equiv T_{*\mu}^{\ \ \mu} = T_*^{\mu\nu}g_{*\mu\nu}, \\
 \alpha (\varphi) &\equiv \frac{d\ln A(\varphi)}{d\varphi}.
\end{align}
It is apparent that $\alpha(\varphi)$ is the only field-dependent
function which couples the scalar field with matter, for
$\alpha(\varphi)=0$ the theory reduces to general relativity.

Finally, the law for energy-momentum conservation
$\tilde{\nabla}_{\nu}\tilde{T}_{\mu}^{\ \nu}=0$ is transformed
into
\begin{equation}
 \nabla_{*\nu}T_{*\mu}^{\ \ \nu} = \alpha (\varphi)T_*\nabla_{*\mu}\varphi, \label{basic3}
\end{equation}
and we set $\varphi_0$ as the cosmological value of the scalar field at infinity.
In this paper, for simplicity, we adopt the same form of conformal
factor $A(\varphi)$ as in Damour and
Esposito-Far\`ese\cite{Damour1993}, which is
\begin{equation}
 A(\varphi) = e^{\frac{1}{2}\beta\varphi^2} \label{A_varphi}
\end{equation}
i.e. $\alpha(\varphi)=\beta \varphi$ where $\beta$ is a real
number. In the case for $\beta=0$, the scalar-tensor theory
reduces to general relativity, while the ``spontaneous
scalarization" occurs for $\beta\le -4.35$ \cite{Harada1998}.

We will model the neutron stars as self-gravitating perfect
fluids, made out of degenerate matter at equilibrium and admitting
cold equation of state. Then the metric describing an unperturbed,
non-rotating, spherically symmetric neutron star can be written as
\begin{equation}
 ds_*^2 = g_{*\mu\nu}dx^{\mu}dx^{\nu}
        = -e^{2\Phi}dt^2 + e^{2\Lambda}dr^2 + r^2(d\theta^2 + \sin^2\theta d\phi^2)
\end{equation}
where
\begin{equation}
 e^{-2\Lambda} = 1-\frac{2\mu(r)}{r},
\end{equation}
while the ``potential'' function $\Phi(r)$ will be calculated later.
The stellar matter is assumed to be a perfect fluid
\begin{equation}
 \tilde{T}_{\mu\nu} = \left(\tilde{\rho} +\tilde{P}\right)\tilde{U}_{\mu}\tilde{U}_{\nu}
                    + \tilde{P}\tilde{g}_{\mu\nu}.
\end{equation}
where $\tilde{U}_{\mu}$  is the four-velocity of the fluid,
$\tilde{\rho}$ is the total energy density in the fluid frame, and
${\tilde p}$ is the pressure.

Spherical symmetry simplifies significantly the procedure of
constructing stellar models. Using equations (\ref{basic1}),
(\ref{basic2}) and  (\ref{basic3}), we can obtain the following
set of equations for the background configuration
\cite{Damour1993,Harada1998}
\begin{align}
 \frac{d\mu}{dr} &= 4\pi G_*r^2A^4\tilde{\rho}+\frac{1}{2}r(r-2\mu)\Psi^2, \label{dmu} \\
 \frac{d\Phi}{dr} &= 4\pi G_*\frac{r^2A^4\tilde{P}}{r-2\mu}+\frac{1}{2}r\Psi^2+\frac{\mu}{r(r-2\mu)},
   \label{dPhi} \\
 \frac{d\varphi}{dr} &= \Psi, \label{dvarphi} \\
 \frac{d\Psi}{dr} &= 4\pi G_*\frac{rA^4}{r-2\mu}
   \left[\alpha(\tilde{\rho}-3\tilde{P})+r(\tilde{\rho}-\tilde{P})\Psi\right]
   -\frac{2(r-\mu)}{r(r-2\mu)}\Psi, \label{dPsi} \\
 \frac{d\tilde{P}}{dr} &= -\left(\tilde{\rho}+\tilde{P}\right)
   \left[\frac{d\Phi}{dr}+\alpha\Psi\right] \nonumber  \\
  &= -\left(\tilde{\rho}+\tilde{P}\right)
   \left[4\pi G_*\frac{r^2A^4\tilde{P}}{r-2\mu}+\frac{1}{2}r\Psi^2+\frac{\mu}{r(r-2\mu)}+\alpha\Psi\right].
   \label{dP}
\end{align}
Near the center the background quantities $\varphi$, $\Psi$,
$\Phi$, ${\tilde P}$ and $\mu$ can be expanded as,
\begin{align}
 \varphi(r)   &= \varphi_c   + \frac{1}{2}\varphi_2 r^2   + O(r^4), \\
 \Psi(r)      &= \varphi '   = \varphi_2 r                + O(r^3), \\
 \Phi(r)      &= \Phi_c      + \frac{1}{2}\Phi_2 r^2      + O(r^4), \\
 \tilde{P}(r) &= \tilde{P}_c + \frac{1}{2}\tilde{P}_2 r^2 + O(r^4), \\
 \mu(r)       &= O(r^3),
\end{align}
where the expansion coefficients are given by
\begin{align}
 \Phi_2      &= 4\pi G_*A^4_c\,\tilde{P}_c, \\
 \varphi_2   &= \frac{4\pi}{3}G_*A^4_c\,\alpha_c\left(\tilde{\rho}_c-3\tilde{P}_c\right), \\
 \tilde{P}_2 &= -\left(\tilde{\rho}_c+\tilde{P}_c\right)\left(\Phi_2+\alpha_c\varphi_2\right),
\end{align}
where $A_c\equiv A(\varphi_c)$ and $\alpha_c \equiv \alpha (\varphi_c)$.

Outside the star, the metric is assumed to be static and
spherically symmetric and since we deal with only one scalar
field, it gets the form  \cite{Damour1992}
\begin{align}
 ds^2_{*} &=-e^{\nu}dt^2+e^{-\nu}d\bar{r}^2+e^{-\nu+\lambda}(d\theta^2+\sin^2\theta d\phi^2), \\
 e^{\nu} &=\left(1-\frac{a}{\bar{r}}\right)^{b/a},\ \ e^{\lambda}=\bar{r}(\bar{r}-a) \label{outer_g},
\end{align}
where $\bar{r}$ is a radial coordinate, given by the following relation
\begin{equation}
  r^2 = \bar{r}^2\left(1-\frac{a}{\bar{r}}\right)^{1-b/a}.
\end{equation}
Moreover, in the above metric (\ref{outer_g}), $a$ and $b$ are some constants, which are connected with the
total scalar charge $\omega_A$ and the total ADM mass
$M_{\mbox{\small ADM}}$ ,i.e., $a^2-b^2=4{\omega_A}^2$ and $b=2M_{\mbox{\small ADM}}$.
Finally, the asymptotic form, at spatial infinity, for the metric and scalar field will be given
as functions of the total ADM mass and the total scalar charge
\begin{align}
  g_{*\mu\nu} &=\eta_{\mu\nu} + \frac{2M_{\mbox{\small ADM}}}{r}\delta_{\mu\nu} + O\left(\frac{1}{r^2}\right), \\
  \varphi &= \varphi_0 + \frac{\omega_A}{r} +O\left(\frac{1}{r^2}\right),
\end{align}
where $\eta_{\mu\nu}$ is the Minkowskian metric.

By matching this exterior solution and the interior metric, one
gets the following relations \cite{Damour1993}
\begin{align}
 M_{\mbox{\small ADM}} &=\frac{R^2\Phi'_s}{G_*}\left(1-\frac{2\mu_s}{R}\right)^{1/2}\exp\left[
      -\frac{\Phi'_s}{\sqrt{(\Phi'_s)^{2}+\Psi_s^2}}\mbox{arctanh}
      \left(\frac{\sqrt{(\Phi'_s)^2+\Psi_s^2}}{\Phi'_s+1/R}\right)\right], \\
 \varphi_{0} &=\varphi_s+\frac{\Psi_s}{\sqrt{(\Phi'_s)^{2}+\Psi_s^2}}
    \mbox{arctanh}\left[\frac{\sqrt{(\Phi'_s)^2+\Psi_s^2}}{\Phi'_s+1/R}\right], \\
 \Phi'_s &= \frac{1}{2}R\Psi_s^2+\frac{\mu_s}{R(R-2\mu_s)},
\end{align}
where the functions with the subscript $s$ correspond to their
actual values at the stellar surface and the prime denotes the
derivative with respect to $r$. 
A more general scheme for fixing the asymptotic value of $\varpi_0$ can be derived using the cosmological model of Damour and Nordtvedt\cite{DN1993}, see for example \cite{Comer1998}. Still, this approach has been applied for $\beta>0$ which is not the subject of our present study.

In order to determine the stellar properties,
an additional equation is needed, i.e. the equation of state (EOS).
Here we will use a polytropic one given by
\begin{align}
 \tilde{P} &= Kn_0m_b\frac{\tilde{n}}{n_0}, \\
 \tilde{\rho} &= \tilde{n}m_b+\frac{\tilde{P}}{\Gamma-1}, \\
 m_b &= 1.66\times 10^{-24}\ \mbox{g}, \\
 n_0 &= 0.1\ \mbox{fm}^{-3}.
\end{align}
We have selected $\Gamma=2.46$ and $K=0.00936$ in agreement with a
fitting to tabulated data for EOS A \cite{Arnett1977}, and
$\Gamma=2.34$ and $K=0.0195$ to fit EOS II \cite{Diaz1985}. In
other words the parameters $\Gamma$ and $K$ are adjusted to fit
the tabulated data from these two realistic equations of state
which have been used in earlier studies of the problem
\cite{Damour1993,Harada1998}. The present study will be
constrained to these two EOS since our aim is to demonstrate the
effect of the scalar field in the neutron star oscillation spectra
while more detailed analysis for a wide range of EOS is underway.

\subsection{Neutron Star Models}
\label{sec:stellar-model}

Here we present some typical stellar models for the two equations
of state under discussion i.e. EOS A and EOS II. For the
construction of the stellar models, there exist three freely
specifiable parameters, these are: the constant $\beta$ of the
conformal factor, the value $\varphi_0$ of the scalar field
$\varphi$ at infinity, and the central density $\tilde{\rho}_c$.
We will also consider positive values of $\varphi_0$, because the basic
equations (\ref{dmu}) -- (\ref{dP}) are symmetric under the
reflection $\varphi\to -\varphi$. Also, we will only consider
stellar models with $\beta=-6$ and $\beta=-8$, for these range of
values the effect of the scalar field is more pronounced. 
Binary pulsar data already suggest larger values for $\beta$ i.e. $\beta \agt -4.5$, 
but this work is based on a different framework and the results 
can be used as an alternative way of constraining the appearence 
of ``sponataneous scalarization" in neutron stars.  

\begin{center}
\begin{figure}[htbp]
\includegraphics[height=17cm,clip]{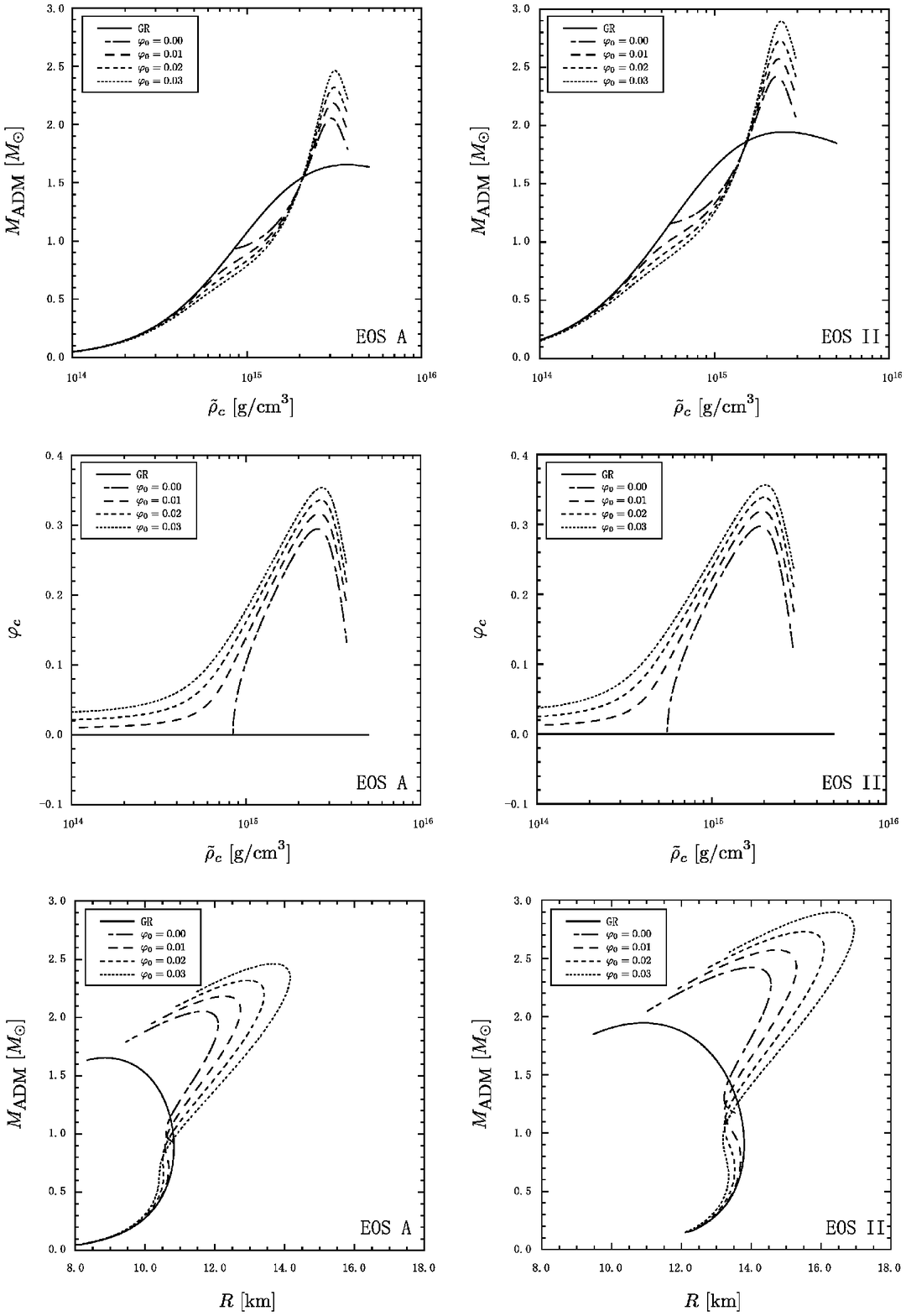}
\caption{
Stellar models with $\beta=-6$ are shown. The left column
corresponds to models for the EOS A while in the right to models
for EOS II. In the upper panels the ADM mass $M_{\mbox{\small
ADM}}$ and the value of the scalar field in the center of the
star, $\varphi_c$ are plotted as functions of the central density.
The effect of the scalar field is apparent for $\rho_c \ge 5\times
10^{14} {\rm gr/cm}^3$. The lower panel is a mass-radius diagram
where the dramatic effect of the scalar field can be observed. In
every panel, the general relativistic stellar models are shown by
a solid line (GR) }
\label{fig_property_beta6}
\end{figure}
\end{center}

In Figure \ref{fig_property_beta6} we show stellar models with
$\beta=-6$. In the left column we plot models for the EOS A and in
the right column models for the EOS II. In the two upper panels we
plot the ADM mass $M_{\mbox{\small ADM}}$ as functions of the
central density, in the two middle panels we plot the value of the
scalar field in the center of the star, $\varphi_c$ as functions
of the central density. It is apparent that the effect of the
scalar field is pronounced for central densities higher than
$5\times 10^{14} {\rm gr/cm}^3$, for lower densities there is no
way that one can trace the existence of a scalar field. The lower
panel is a ``typical" mass-radius relation where one can observe
the dramatic effect of the scalar field in the stellar structure.
For comparison in every panel the general relativistic stellar
models are shown by a solid line (GR)

\begin{center}
\begin{figure}[htbp]
\includegraphics[height=17cm,clip]{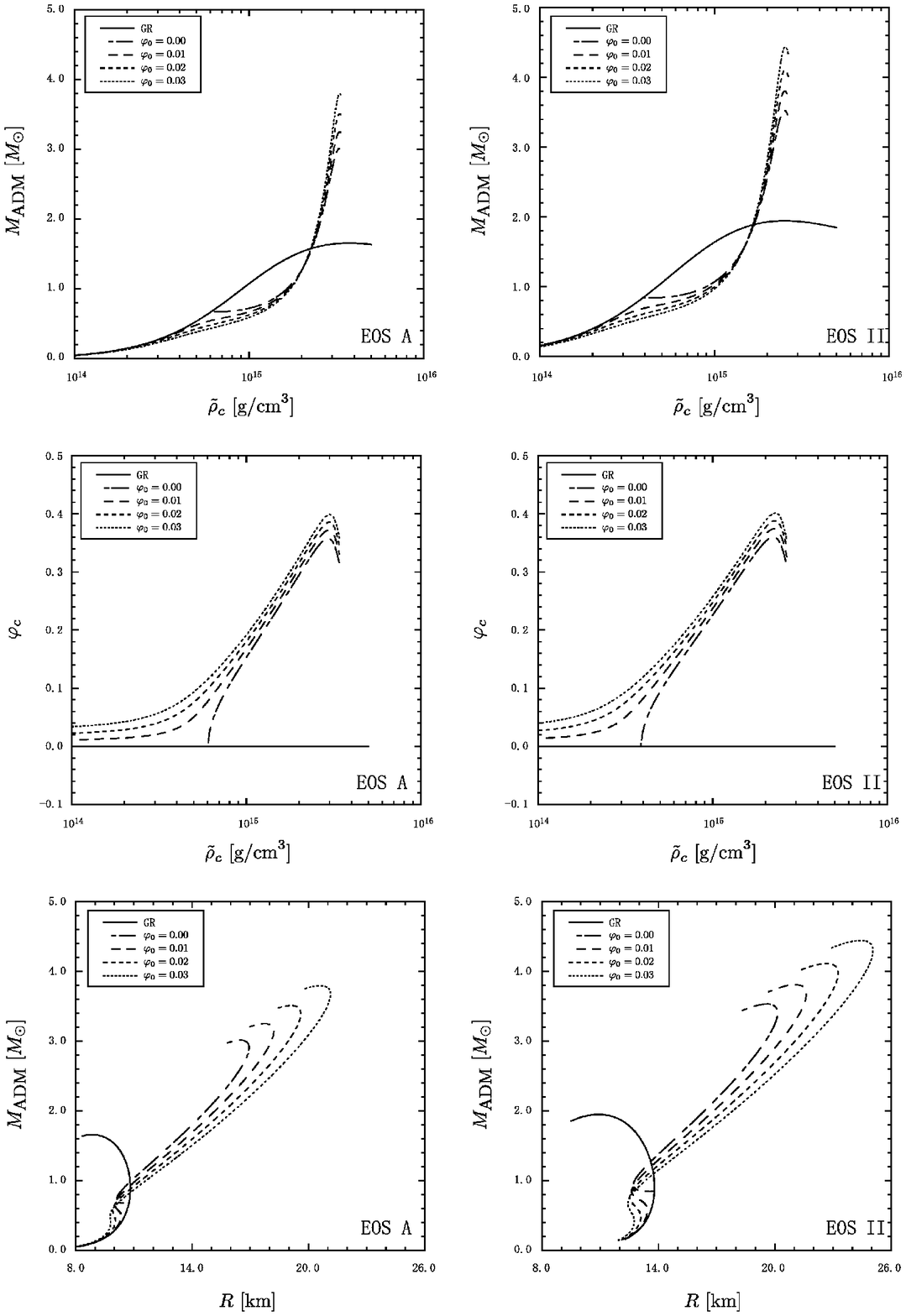}
\caption{
Stellar models with $\beta=-8$ are shown. The left column
corresponds to models for the EOS A while in the right to models
for EOS II. In the upper panels the ADM mass $M_{\mbox{\small
ADM}}$ and the value of the scalar field in the center of the
star, $\varphi_c$ are plotted as functions of the central density.
The effect of the scalar field is apparent for  $\rho_c \ge
3\times 10^{14} {\rm gr/cm}^3$. The lower panel is a mass-radius
diagram where the dramatic effect of the scalar field can be
observed. In every panel, the general relativistic stellar models
are shown by a solid line (GR) }
\label{fig_property_beta8}
\end{figure}
\end{center}

In Figure \ref{fig_property_beta8} we present stellar models for
$\beta=-8$, the ordering of the graphs is as in Figure
\ref{fig_property_beta6}. Now the effect of the scalar field is
more pronounced than before. Its presence can be traced for even lower
densities while for higher densities the stellar models divert
considerable from those of GR. The effects on the mass and radius
are even more dramatic in the mass-radius diagrams in the lower
panels.

Finally, in order to stress the effect of the scalar field  on the
maximum mass of neutron stars we show in two tables the relevant
parameters for each EOS. The maximum masses are well beyond the
observed values of neutron star. This is also true for the
observed redshifts. It is worth mentioning that the redshift of
the maximum mass models is quite unaffected by the changes in the
central density and the scalar field parameter. Although for these
maximum mass models the redshift is unusually large, for the
typical models that we will use the surface redshift $z$ is below the
maximum observed redshift which is $z=0.35$ \cite{Cottam2002}.
This means that neutron stars of the scalar-tensor theories cannot
be exclude by the present electromagnetic observations.

\begin{table}[htbp]
      \begin{center}
          \leavevmode
          \caption{
          The stellar parameters for models with maximum ADM mass (EOS A).
          }
          \begin{tabular}{cccccccccc}
            \hline\hline
              & $\beta$ & $\varphi_0$ & $M_{\mbox{\small ADM}}/M_{\odot}$ & $\tilde{\rho}_c$ (g/cm$^3$) &
              $R$ (km) & $\varphi_c$ & $M_{\mbox{\small ADM}}/R$ &z&\\ \hline
              &  0.00 & ---  & 1.655 & $3.742\times 10^{15}$ & 8.56  & ---    & 0.285& 0.527&\\
              & -6.00 & 0.00 & 2.054 & $3.048\times 10^{15}$ & 11.64 & 0.2676 & 0.261& 0.445&\\
              & -6.00 & 0.03 & 2.462 & $3.199\times 10^{15}$ & 13.67 & 0.3269 & 0.266& 0.462&\\
              & -8.00 & 0.00 & 3.010 & $3.294\times 10^{15}$ & 16.49 & 0.3338 & 0.2695& 0.473&\\
              & -8.00 & 0.03 & 3.792 & $3.317\times 10^{15}$ & 20.61 & 0.3755 & 0.2717& 0.480&\\
            \hline\hline
          \end{tabular}
          \label{table:property_EOSA}
      \end{center}
\end{table}
\begin{table}[htbp]
      \begin{center}
          \leavevmode
          \caption{
       The stellar parameters for models with maximum ADM mass (EOS II).
          }
          \begin{tabular}{cccccccccc}
            \hline\hline
              & $\beta$ & $\varphi_0$ & $M_{\mbox{\small ADM}}/M_{\odot}$ & $\tilde{\rho}_c$ (g/cm$^3$) &
              $R$ (km) & $\varphi_c$ & $M_{\mbox{\small ADM}}/R$ & z&\\ \hline
              &  0.00 & ---  & 1.946 & $2.543\times 10^{15}$ & 10.90 & ---    & 0.264& 0.454&\\
              & -6.00 & 0.00 & 2.423 & $2.282\times 10^{15}$ & 14.00 & 0.2724 & 0.256& 0.430&\\
              & -6.00 & 0.03 & 2.896 & $2.423\times 10^{15}$ & 16.32 & 0.3307 & 0.262& 0.450&\\
              & -8.00 & 0.00 & 3.529 & $2.523\times 10^{15}$ & 19.54 & 0.3366 & 0.267& 0.464&\\
              & -8.00 & 0.03 & 4.442 & $2.543\times 10^{15}$ & 24.42 & 0.3792 & 0.269& 0.470&\\
            \hline\hline
          \end{tabular}
          \label{table:property_EOSII}
      \end{center}
\end{table}

%
%
%

\section{Perturbation Equations in the Cowling approximation}
\label{sec:III}

In this section we present the perturbation equations for
non-radial oscillations of spherically symmetric neutron stars in
scalar-tensor theory. We derive the equations, describe the
numerical method for getting the spectra and finally we present
results for a few characteristic values of the scalar field
$\beta$. The presentation in this section is quite extended, we
have chosen to show all the details of the calculation since this
is the first article dealing with the stellar perturbations in
scalar-tensor theory and there is a wealth of new functions and
notations. We will also keep arbitrary the functional form of the
scalar function $\alpha(\varphi)$ which will be fixed only later
during the numerical calculations. In this first investigation we
will not consider the full problem but instead we will restrict it
to the so called ``Cowling approximation''. This means that the
fluid is perturbed on a fixed background, in this way we freeze
the spacetime and scalar field perturbations i.e. $\delta
\tilde{g}_{\mu\nu}=0$ and $\delta\varphi=0$. 
The Cowling approximation limits our study to the modes which are
directly related
to the fluid perturbations i.e the $f$, $p$ and $g$-modes, while
we cannot study the emission of gravitational and scalar waves
neither the families of the spacetime \cite{Kokkotas1992} and
scalar modes.

The fluid perturbations will be described by the Lagrangian displacement vector
\begin{equation}
\tilde{\xi}^{i}=(\tilde{\xi}^r,\tilde{\xi}^{\theta},\tilde{\xi}^{\phi})=\left(e^{-\Lambda}W,-V\frac{\partial}{\partial\theta},-\frac{1}{\sin^2\theta}V\frac{\partial}{\partial\phi}\right)\frac{1}{r^2}Y_{lm}
\end{equation}
where $W$ and $V$ are functions with of $t$ and $r$.
The unperturbed 4-velocity $\tilde{U}^{(B)\mu}$ is
\begin{equation}
 \tilde{U}^{(B)\mu} = \left(\frac{1}{A(\varphi) e^{\Phi}},0,0,0\right),
\end{equation}
hereafter an index ``$(B)$" will denote the unperturbed
quantities. The perturbed 4-velocity $\delta\tilde{U}^{\mu}$ has
the form
\begin{align}
 \delta\tilde{U}^{\mu}
   = \left(0,e^{-\Lambda}\frac{\partial W}{\partial t},
            -\frac{\partial V}{\partial t}\frac{\partial}{\partial\theta},
            -\frac{1}{\sin^2\theta}\frac{\partial V}{\partial t}\frac{\partial}{\partial\phi}\right)\frac{1}{A(\varphi)r^2}e^{-\Phi}Y_{lm}.
\end{align}
The perturbed energy-momentum tensor $\delta\tilde{T}_{\mu}^{\
\nu}$ in the Cowling approximation gets the form
\begin{align}
 \delta\tilde{T}_{\mu}^{\ \nu} &= (\delta\tilde{\rho}+\delta\tilde{P})
    \tilde{g}^{(B)}_{\mu\beta}\tilde{U}^{(B)\beta}\tilde{U}^{(B)\nu} \nonumber \\
    & +(\tilde{\rho}^{(B)}+\tilde{P}^{(B)})\tilde{g}^{(B)}_{\mu\beta}
    (\tilde{U}^{(B)\beta}\delta\tilde{U}^{\nu}+\delta\tilde{U}^{\beta}\tilde{U}^{(B)\nu})
    +\delta\tilde{P}\,\delta_{\mu}^{\ \nu}.
\end{align}
The Lagrangian variation of the baryon number density $\Delta\tilde{n}$ is
\begin{equation}
 \frac{\Delta\tilde{n}}{\tilde{n}}
   = -\tilde{\nabla}^{(3)}_k \tilde{\xi}^{k}-\frac{\delta\tilde{g}}{2\tilde{g}^{(B)}}, \label{deltan}
\end{equation}
where $\tilde{\nabla}^{(3)}_k$  denotes the covariant derivative
in a 3-dimensional with  metric $\tilde{g}_{\mu\nu}$ and $\Delta$
the Lagrangian variation. In this equation, the first term of the
right hand side expresses the 3-dimensional divergence of the
fluid, and the second term is the amount of the volume change due
to the metric perturbation, but because we use the Cowling
approximation we neglect this second term. By employing the
above perturbation variables, equation (\ref{deltan}) for the
Lagrangian variation of the baryon number density will be written
as
\begin{equation}
 \frac{\Delta\tilde{n}}{\tilde{n}}
      = -\left[\frac{1}{r^2}e^{-\Lambda}\frac{\partial W}{\partial r}+\frac{l(l+1)V}{r^2}
      +3\alpha\Psi e^{-\Lambda}\frac{W}{r^2}\right]Y_{lm}. \label{A}
\end{equation}
For adiabatic perturbations, using the first law of
thermodynamics, we can get the following relation between the
change of the baryon number density and the pressure
\begin{equation}
 \Delta \tilde{\rho} = \frac{\tilde{\rho}+\tilde{P}}{\tilde{n}}\Delta\tilde{n}.  \label{1}
\end{equation}
Therefore we can express the Eulerian density variation $\delta\tilde{\rho}$ as,
\begin{equation}
 \delta\tilde{\rho} = (\tilde{\rho}^{(B)}+\tilde{P}^{(B)})\frac{\Delta\tilde{n}}{\tilde{n}}
    -\frac{\partial\tilde{\rho}^{(B)}}{\partial r}e^{-\Lambda}\frac{W}{r^2}Y_{lm}, \label{B}
\end{equation}
here we have used the relation between the Lagrangian perturbation
$\Delta$ and Eulerian perturbation $\delta$
\begin{equation}
 \Delta\tilde{\rho}(t,r) = \tilde{\rho}(t,r+\tilde{\xi}^r)-\tilde{\rho}^{(B)}(t,r)\simeq
   \delta\tilde{\rho}+\frac{\partial \tilde{\rho}^{(B)}}{\partial r}\tilde{\xi}^r.
\end{equation}
From the definition of the adiabatic constant
\begin{equation}
 \gamma\equiv \left(\frac{\partial\ln\tilde{P}}{\partial\ln\tilde{n}}\right)_s
    =\frac{\tilde{n}\Delta\tilde{P}}{\tilde{P}\Delta\tilde{n}}, \label{2}
\end{equation}
we can derive the form of the Eulerian variation of the pressure
\begin{equation}
 \delta\tilde{P} = \gamma\tilde{P}^{(B)}\frac{\Delta\tilde{n}}{\tilde{n}}
    -\frac{\partial\tilde{P}^{(B)}}{\partial r}e^{-\Lambda}\frac{W}{r^2}Y_{lm}.
\end{equation}
Finally, the combination of equations (\ref{1}) and (\ref{2}),
provides the standard form of the adiabatic constant
\begin{equation}
 \gamma = \frac{\tilde{\rho}+\tilde{P}}{\tilde{P}}\left(\frac{\partial\tilde{P}}{\partial\tilde{\rho}}\right)_s.
\end{equation}

By taking a variation of the equation for the  conservation of
energy-momentum $\tilde\nabla_{\nu}\tilde{T}_{\mu}^{\ \nu}=0$, we
can get the equation describing the perturbations of the fluid in
the Cowling approximation
\begin{equation}
 \delta\tilde{T}_{\mu\ ,\nu}^{\ \nu}-\Gamma_{*\ \mu\nu}^{\ \alpha}\delta\tilde{T}_{\alpha}^{\ \nu}
    +\Gamma_{*\ \alpha\nu}^{\ \nu}\delta\tilde{T}_{\mu}^{\ \alpha}
    -\alpha\varphi_{,\mu}\delta\tilde{T}_{\nu}^{\ \nu}+4\alpha\varphi_{,\nu}\delta\tilde{T}_{\mu}^{\ \nu}=0.
  \label{D}
\end{equation}
The analytic form of the above equation for the values of the index $\mu=1,2$  is :
\begin{align}
 \tilde\nabla_{\nu}\delta\tilde{T}_1^{\ \nu}&=0 \nonumber \\
 \Leftrightarrow\ \ &(\tilde{\rho}+\tilde{P})\frac{1}{r^2}e^{\Lambda-2\Phi}\frac{\partial^2 W}{\partial t^2}
    \nonumber \\
    &-\frac{\partial}{\partial r}\left[\gamma\tilde{P}\left\{\frac{1}{r^2}e^{-\Lambda}\frac{\partial W}{\partial r}
    +\frac{l(l+1)V}{r^2}+3\alpha\Psi e^{-\Lambda}\frac{W}{r^2}\right\}\right]
    -\frac{\partial}{\partial r}\left(e^{-\Lambda}\frac{d\tilde{P}}{dr}\frac{W}{r^2}\right) \nonumber \\
    &+\frac{d\tilde{P}}{dr}
    \left[\frac{1}{r^2}e^{-\Lambda}\frac{\partial W}{\partial r}
    +\frac{l(l+1)V}{r^2}+3\alpha\Psi e^{-\Lambda}\frac{W}{r^2}\right]
    -\left(\frac{d\Phi}{dr}+\alpha\Psi\right)
    e^{-\Lambda}\frac{d\tilde{\rho}}{dr}\frac{W}{r^2} \nonumber \\
    &+\frac{d\tilde{P}}{dr}\frac{d\tilde{P}}{d\tilde{\rho}}
    \left[\frac{1}{r^2}e^{-\Lambda}\frac{\partial W}{\partial r}
    +\frac{l(l+1)V}{r^2}+3\alpha\Psi e^{-\Lambda}\frac{W}{r^2}\right]
    -\left(\frac{d\Phi}{dr}+\alpha\Psi\right)
    e^{-\Lambda}\frac{d\tilde{P}}{dr}\frac{W}{r^2}=0. \label{F} \\
 \tilde\nabla_{\nu}\delta\tilde{T}_2^{\ \nu}&=0  \nonumber \\
 \Leftrightarrow\ \ &(\tilde{\rho}+\tilde{P})e^{-2\Phi}\frac{\partial^2 V}{\partial t^2} 
    +\gamma\tilde{P}\left[\frac{1}{r^2}e^{-\Lambda}\frac{\partial W}{\partial r}
    +\frac{l(l+1)V}{r^2}+3\alpha\Psi e^{-\Lambda}\frac{W}{r^2}\right]
    +e^{-\Lambda}\frac{d\tilde{P}}{dr}\frac{W}{r^2}=0. \label{G}
\end{align}
By assuming a harmonic dependence on time the perturbation
functions will be written as  $W(t,r)=W(r)e^{i\omega t}$ and
$V=V(t,r)=V(r)e^{i\omega t}$ and the above system of equations
gets the form
\begin{align}
 &\omega^2(\tilde{\rho}+\tilde{P})e^{\Lambda-2\Phi}\frac{W}{r^2}
    +\frac{d}{dr}\left[\gamma\tilde{P}\left\{\frac{1}{r^2}e^{-\Lambda}\frac{dW}{dr}+\frac{l(l+1)V}{r^2}
    +3\alpha\Psi e^{-\Lambda}\frac{W}{r^2}\right\}\right]
    +\frac{d}{dr}\left(e^{-\Lambda}\frac{d\tilde{P}}{dr}\frac{W}{r^2}\right) \nonumber \\
    &\hspace{1cm}-\frac{d\tilde{P}}{dr}\left[\frac{1}{r^2}e^{-\Lambda}\frac{dW}{dr}
    +\frac{l(l+1)V}{r^2}+3\alpha\Psi e^{-\Lambda}\frac{W}{r^2}\right]
    +\left(\frac{d\Phi}{dr}+\alpha\Psi\right)
    e^{-\Lambda}\frac{d\tilde{\rho}}{dr}\frac{W}{r^2} \nonumber \\
    &\hspace{1cm}-\frac{d\tilde{P}}{dr}\frac{d\tilde{P}}{d\tilde{\rho}}
    \left[\frac{1}{r^2}e^{-\Lambda}\frac{dW}{dr}
    +\frac{l(l+1)V}{r^2}+3\alpha\Psi e^{-\Lambda}\frac{W}{r^2}\right]
    +\left(\frac{d\Phi}{dr}+\alpha\Psi\right)
    e^{-\Lambda}\frac{d\tilde{P}}{dr}\frac{W}{r^2}=0,
    \label{F1} \\
 -&\omega^2 e^{-2\Phi}(\tilde{\rho}+\tilde{P})V+\frac{\gamma\tilde{P}}{r^2}\left[e^{-\Lambda}\frac{dW}{dr}
    +l(l+1)V+3\alpha\Psi e^{-\Lambda}W\right]+e^{-\Lambda}\frac{d\tilde{P}}{dr}\frac{W}{r^2}=0. \label{G1}
\end{align}
A further simplification, can be achieved, for the first of the
above equations by using an appropriate combination of the form
$d(\ref{G1})/dr-(\ref{F1})$, this leads to the equation
\begin{align}
 -\omega^2&\frac{d}{dr}\left[e^{-2\Phi}(\tilde{\rho}+\tilde{P})V\right]
    -\omega^2(\tilde{\rho}+\tilde{P})e^{\Lambda-2\Phi}\frac{W}{r^2} \nonumber \\
    &+\frac{d\tilde{P}}{dr}
    \left[\frac{1}{r^2}e^{-\Lambda}\frac{\partial W}{\partial r}
    +\frac{l(l+1)V}{r^2}+3\alpha\Psi e^{-\Lambda}\frac{W}{r^2}\right]
    -\left(\frac{d\Phi}{dr}+\alpha\Psi\right)
    e^{-\Lambda}\frac{d\tilde{\rho}}{dr}\frac{W}{r^2} \nonumber \\
    &+\frac{d\tilde{P}}{dr}\frac{d\tilde{P}}{d\tilde{\rho}}
    \left[\frac{1}{r^2}e^{-\Lambda}\frac{\partial W}{\partial r}
    +\frac{l(l+1)V}{r^2}+3\alpha\Psi e^{-\Lambda}\frac{W}{r^2}\right]
    -\left(\frac{d\Phi}{dr}+\alpha\Psi\right)
    e^{-\Lambda}\frac{d\tilde{P}}{dr}\frac{W}{r^2}=0.
\end{align}
which can be further simplified by proper substitutions of
$d{\tilde P}/dr$  and $dW/dr$  from equations (\ref{dP}) and
(\ref{G1})
\begin{equation}
 \frac{dV}{dr} = 2\frac{d\Phi}{dr}V-e^{\Lambda}\frac{W}{r^2}. \label{F2}
\end{equation}
Thus from equations (\ref{G1}) and (\ref{F2}), we get the
following simple system of couple ODEs  \begin{align}
 \frac{dW}{dr} &= \frac{d\tilde{\rho}}{d\tilde{P}}
    \left[\omega^2r^2e^{\Lambda-2\Phi}V+\left(\frac{d\Phi}{dr}+\alpha\Psi\right)W\right]
    -l(l+1)e^{\Lambda}V-3\alpha\Psi W, \label{*1} \\
 \frac{dV}{dr} &= 2\frac{d\Phi}{dr}V-e^{\Lambda}\frac{W}{r^2}, \label{*2}
\end{align}
which together with the appropriate boundary conditions at the
center and at the surface constitute an eigenvalue problem for the
parameter $\omega$ (the eigenfrequency).

The above system of perturbation equations (\ref{*1}) and
(\ref{*2}) near the stellar center gets the following simple form
\begin{align}
 \frac{dW}{dr}&+l(l+1)V\approx 0, \label{*11} \\
 \frac{dV}{dr}&+\frac{W}{r^2}\approx 0. \label{*21}
\end{align}
with the following set of approximate normal solutions
\begin{align}
 W(r) &= Br^{l+1}+\cdots, \label{W0} \\
 V(r) &= -\frac{B}{l}r^l+\cdots. \label{V0}
\end{align}
where $B$ is an arbitrary constant. The two approximate solutions
near the center, Eq. (\ref{W0}) and (\ref{V0}), suggest the introduction
of two new perturbation functions, $W(r)\equiv\bar{W}(r)r^{l+1}$
and $V(r)\equiv\bar{V}(r)r^{l}$. After this change, the
approximate solutions near the center become
\begin{align}
 \bar{W}(r) &= \bar{W}_c+\frac{1}{2}\bar{W}_2r^2+\cdots, \\
 \bar{V}(r) &= \bar{V}_c+\frac{1}{2}\bar{V}_2r^2+\cdots,
\end{align}
where $\bar{W}_c=B$, $\bar{V}_c=-B/l$ and the coefficients
$\bar{W}_2$ and $\bar{V}_2$ of the second order terms are
\begin{align}
 \bar{V}_2 &= \frac{1}{2l+3}\left[3\alpha_c\varphi_2\bar{W}_c+2(l+3)\Phi_2\bar{V}_c
    - \frac{\tilde{\rho}_2}{\tilde{P}_2}\left\{\omega^2e^{-2\Phi_c}\bar{V}_c
    + (\Phi_2+\alpha_c\varphi_2)\bar{W}_c\right\}\right], \\
 \bar{W}_2 &= 4\Phi_2\bar{V}_c-(l+2)\bar{V}_2.
\end{align}

At the surface, the boundary condition is the vanishing of the
Lagrangian perturbation of the pressure $\Delta\tilde{P}=0$. The
Lagrangian perturbation of the pressure is described by
$\Delta\tilde{P}=\gamma\tilde{P}\Delta\tilde{n}/\tilde{n}$, and
thus making use of equation (\ref{2}), we get the following
boundary condition at $r=R$
\begin{equation}
 (\tilde{\rho}+\tilde{P})\left[\omega^2e^{-2\Phi}V+e^{-\Lambda}\left(\frac{d\Phi}{dr}+\alpha\Psi\right)
    \frac{W}{r^2}\right]=0.
\end{equation}
For this new functions the system of perturbation equations becomes
\begin{align}
\label{eq:final1}
 \frac{d\bar{W}}{dr} &= \frac{d\tilde{\rho}}{d\tilde{P}}\left[\omega^2e^{\Lambda-2\Phi}r\bar{V}
    +\left(\frac{d\Phi}{dr}+\alpha\Psi\right)\bar{W}\right]-\frac{l(l+1)}{r}e^{\Lambda}\bar{V}
    -\left(\frac{l+1}{r}+3\alpha\Psi\right)\bar{W}, \\
 \frac{d\bar{V}}{dr} &= \left(2\frac{d\Phi}{dr}-\frac{l}{r}\right)\bar{V}-e^{\Lambda}\frac{\bar{W}}{r},
\label{eq:final2}
\end{align}
with following boundary conditions at the center and the surface
\begin{align}
\label{eq:bc1}
 \bar{W}=-l\bar{V}\ \ \ \ ({\rm at}\ r=0), \\
 \omega^2e^{-2\Phi}r\bar{V}+e^{-\Lambda}\left(\frac{d\Phi}{dr}+\alpha\Psi\right)\bar{W}=0
    \ \ \ \ ({\rm at}\ r=R). \label{eq:bc2}
\end{align}
Equations (\ref{eq:final1}), (\ref{eq:final2}), (\ref{eq:bc1}) and
(\ref{eq:bc2}) form a well posed eigenvalue problem for the real
eigenfrequency $\omega^2$. Using shooting method we can get in a
quite simple way the eigenvalues i.e. the characteristic
frequencies of the oscillations. It should notice that the above
system of perturbation equations and boundary conditions has been
derived for an arbitrary form of the scalar function
$\alpha(\varphi)$, later in the numerical calculations we will fix
its functional form to the one of equation $\alpha(\varphi)$.

\begin{center}
\begin{figure}[htbp]
\includegraphics[height=8cm,clip]{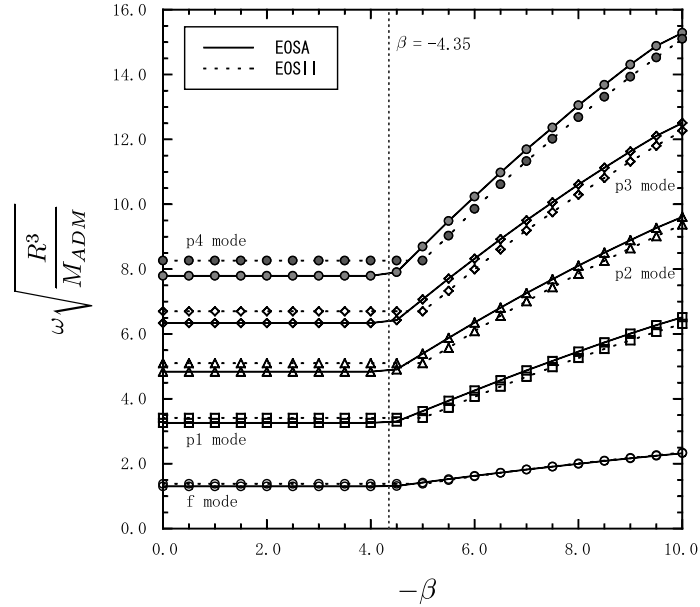}
\caption{
The normalized eigenvalues $\omega$ for the first few modes ($f$,
$p_1$, $p_2$, $p_3$ and $p_4$) are shown as functions the
parameter $\beta$ for the equations of state EOS A (solid line)
and EOS II (dashed line). The effect of the ``spontaneous
scalarization'' is more pronounced for the higher modes. The
asymptotic values of the scalar field and of ADM mass are fixed to
the values $\varphi_0=0.0$ and $M_{ADM}=1.4M_{\odot}$. }
\label{fig_dependence-beta}
\end{figure}
\end{center}
\subsection{The Oscillation Spectra}
\label{sec:spectra}

The spectrum of an oscillating neutron star is directly related to
its parameters, mass, radius and EOS \cite{Andersson1998}. As we
have seen in Section \ref{sec:II} the presence of the scalar field
on the background star is influencing both the mass and the radius,
while in the way that it enters in the equilibrium equations it has a
role of an extra pressure term i.e. it seems to alter the
actual EOS.  As before we will restrict our study only two
equations of state i.e. EOS A and EOS II, and we will use
models from those described in the previous section
\ref{sec:stellar-model}.

First, we will examine the effect of the scalar factor $\beta$ on
the frequency and especially  whether the ``spontaneous
scalarization'' can be traced in the spectrum. A discontinuous
change in a system, as one varies its parameters, signals a
catastrophic behavior and the so called ``spontaneous
scalarization"\cite{Damour1993} is related to it. Harada
\cite{Harada1998} shown that in the scalar-tensor theory the value
$\beta=-4.35$ is the critical one for the ``spontaneous
scalarization''. DeDeo and Psaltis \cite{DeDeo2003} verified that
the effects of the scalar field on the line redshifts of neutron
stars become pronounced for values $\beta \alt-4.35$. We show here
that the spontaneous scalarization is also present in the
oscillation spectra of neutron stars. In order to study this
effect we fixed the asymptotic value of the scalar field to
$\varphi_0=0.0$  and we constructed a sequence of stellar models
with $M=1.4M_\odot$ by varying $\beta$. In Figure
\ref{fig_dependence-beta} the effect of varying $\beta$ on the
frequencies of the fluid becomes immediately apparent. For values
$\beta \alt -4.35$ the frequency gets a sharp change and increases
linearly with decreasing $\beta$ signaling the ``spontaneous
scalarization". The effect becomes more pronounced for the higher
$p$-modes as $\partial\omega/\partial(-\beta)$ increases with the
order of the mode showing the dramatic impact of the ``spontaneous
scalarization'' on the spectrum. By studying the first thirty
modes for each equation of state we have actually found that:
\begin{equation}
\frac{\partial \omega_n}{\partial (-\beta)}\approx \frac{n}{4} \quad \mbox{for}
\quad \beta \alt -4.35
\end{equation}
where $\omega_n$ is the frequency of the $n$th mode. This relation
seems to be independent of the equation of state and suggests that
a possible tracing (via electromagnetic or gravitational
observations) of the higher $p$-modes will signal the existence of
a scalar field even for small deviations from general relativity.

The idea of a gravitational wave asteroseismology
\cite{Andersson1998,Kokkotas2001} was based on empirical relations
that can be drawn for the relation of the stellar parameters to
the eigenfrequencies of an oscillating neutron star. These
empirical relations were derived by taking into account data for a
dozen or more EOS, and  it was shown that through these relations
one can extract the stellar parameters by analyzing the
gravitational wave signal of an oscillating neutron star. The
easiest relation to be understood on intuitive physical grounds is
the one between the fundamental oscillation mode the $f$-mode and
the average density. This relation emerges naturally by combining
the time that a perturbation needs to propagate across the star
and the sound speed, this boils down to a linear relation between
the period of oscillation and the average density or the star or
better $\omega^2\sim M/R^3$. This was the reason that we have
normalized the frequencies in Figure \ref{fig_dependence-beta}
with the average density. According to \cite{Andersson1998} the empirical
relation between the $f$-mode frequency and the average density of
typical neutron stars is
\begin{equation}
f_{\rm{f-mode}} ({\rm kHz})\approx 0.78 +1.63 \left(\frac{M}{1.4M_\odot}\right)^{1/2} \left(\frac{R}{10{\rm km}}\right)^{-3/2}
\label{eq:f-mode}
\end{equation}
%
\begin{figure}[htbp]
\centering
\includegraphics[height=8cm,clip]{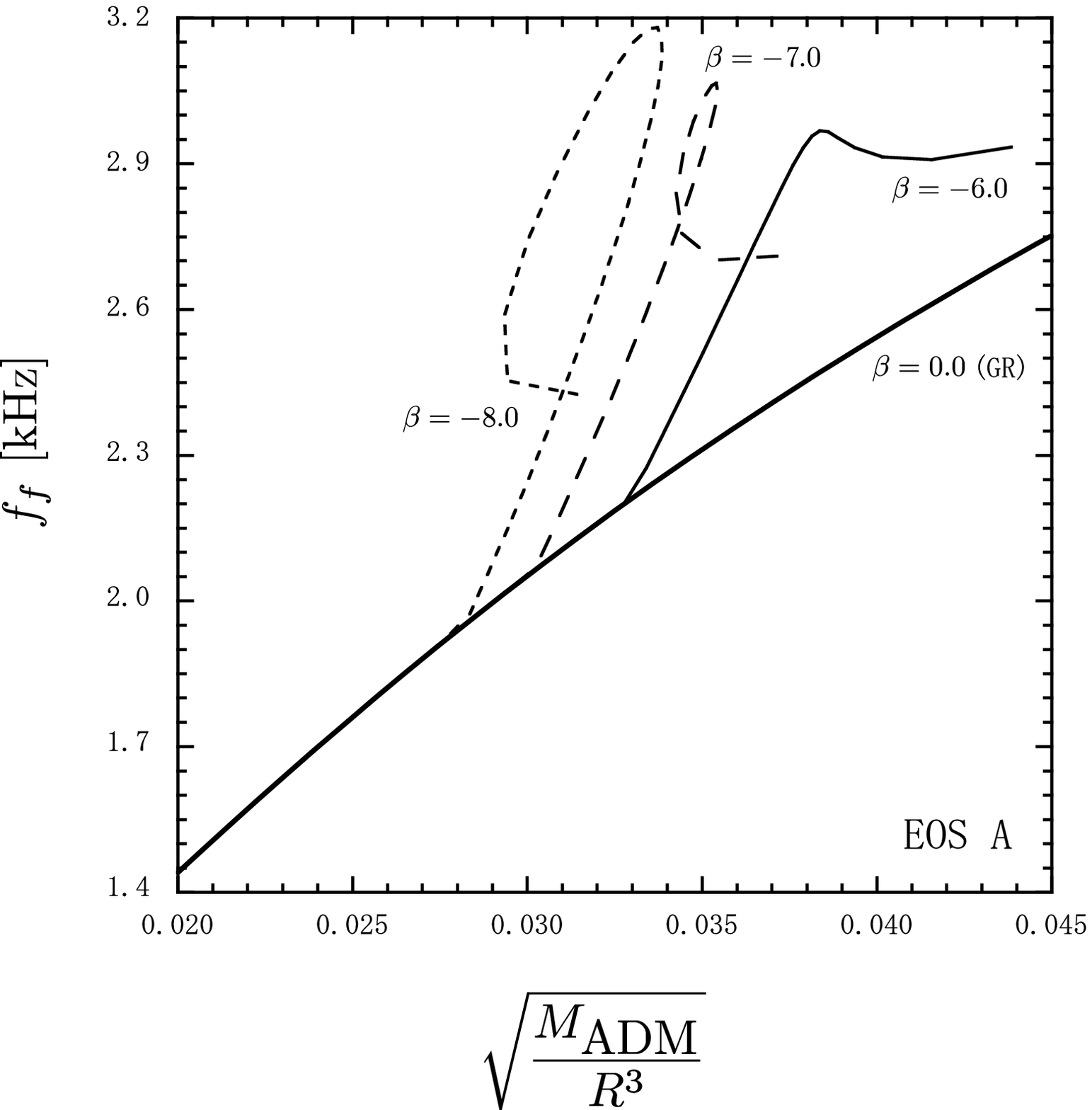}
\includegraphics[height=8cm,clip]{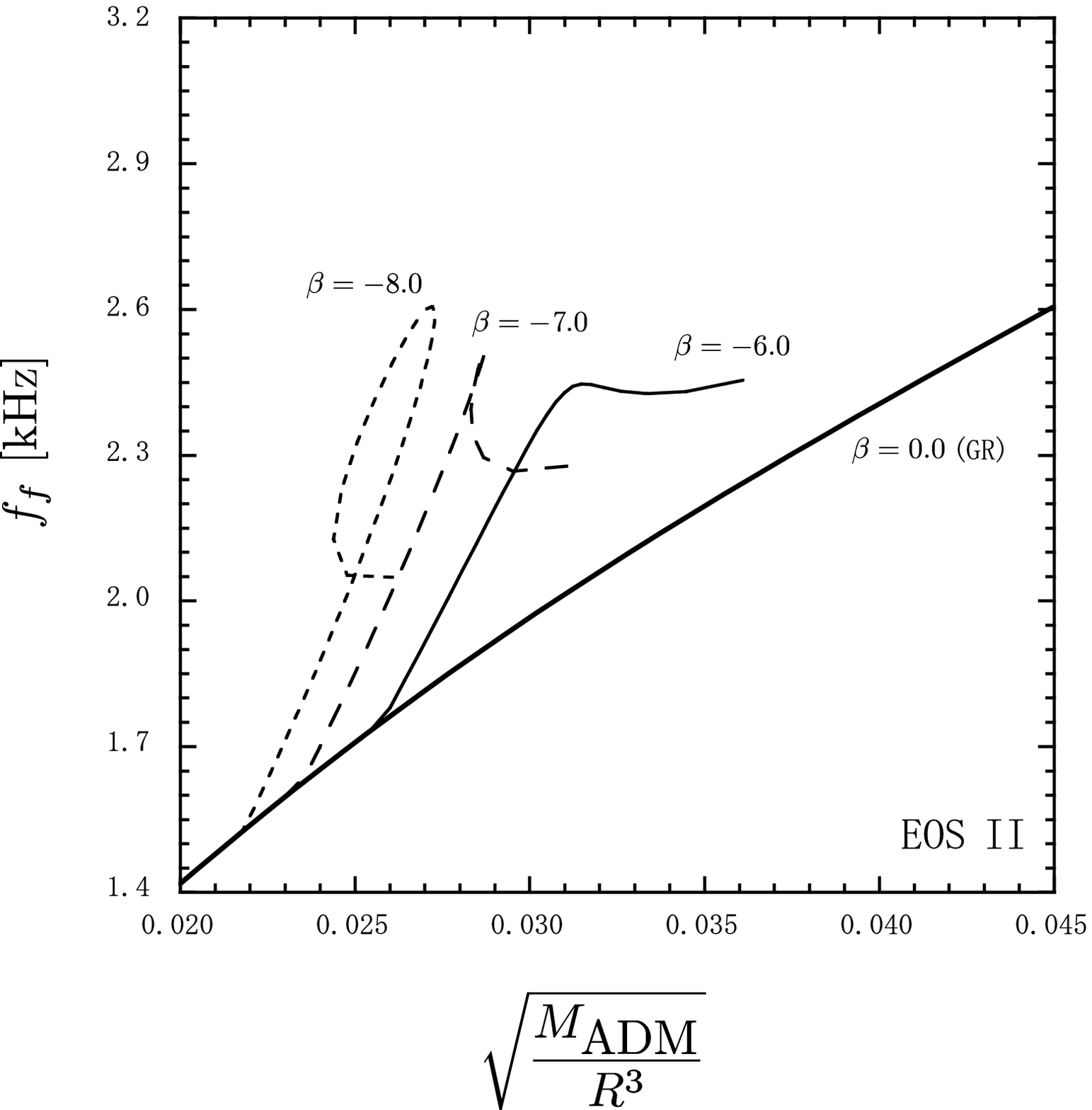}
\caption{ The frequency of the $f$-mode as function of the
averaged density $(M_{\rm ADM}/R^3)^{1/2}$ of the star (notice
that $f_f=\omega_f/2\pi$). The thick solid line corresponds to the
values of the mode for $\beta=0.0$ (GR) while we also show the
effect of the scalar field for three values of the scalar
parameter $\beta$ (-8.0, -7.0 and -6.0). Here we have assumed that
$\varphi_0=0.0$. The left panel corresponds to EOS A and the right
panel to EOS II. } \label{fig:fig4}
\end{figure}
%
Almost, all EOS follow this empirical relation (including the two
EOS used in this article). This observation suggests a unique way
in estimating the average density of a star via its $f$-mode
frequency, and it can be a very good observational test for the
neutron stars in scalar-tensor theories. In figure \ref{fig:fig4}
we draw the frequency as function of the averaged density. The
actual relations of the frequency of the $f$-mode as functions of
the average density for general relativistic neutron stars are
shown as thick solid lines in both panels of figure
\ref{fig:fig4}. Both solid lines correspond to normal neutron
stars with $\beta=0$ and follow the empirical relation
(\ref{eq:f-mode}) with quite small error. The introduction of a
scalar field even in moderate central densities alters completely
the behavior of the $f$-mode for both EOS. The frequencies grow
considerably faster as functions of the average density, for
$\beta \alt -4.35$, and the three examples that we have chosen
($\beta=-6.0$, -7.0 and -8.0) show exactly this behavior. The
change is quite dramatic even for typical neutron stars with
average density. Depending on the value of the parameter $\beta$
they become $30-50\%$ larger than those of a general relativistic
neutron star.  This can be an observable effect, since the
detection of frequencies are higher than those expected from a
typical neutron star, would signal in a unique way the presence of
a scalar field.
\begin{figure}[htbp]
\centering
\includegraphics[height=7cm,clip]{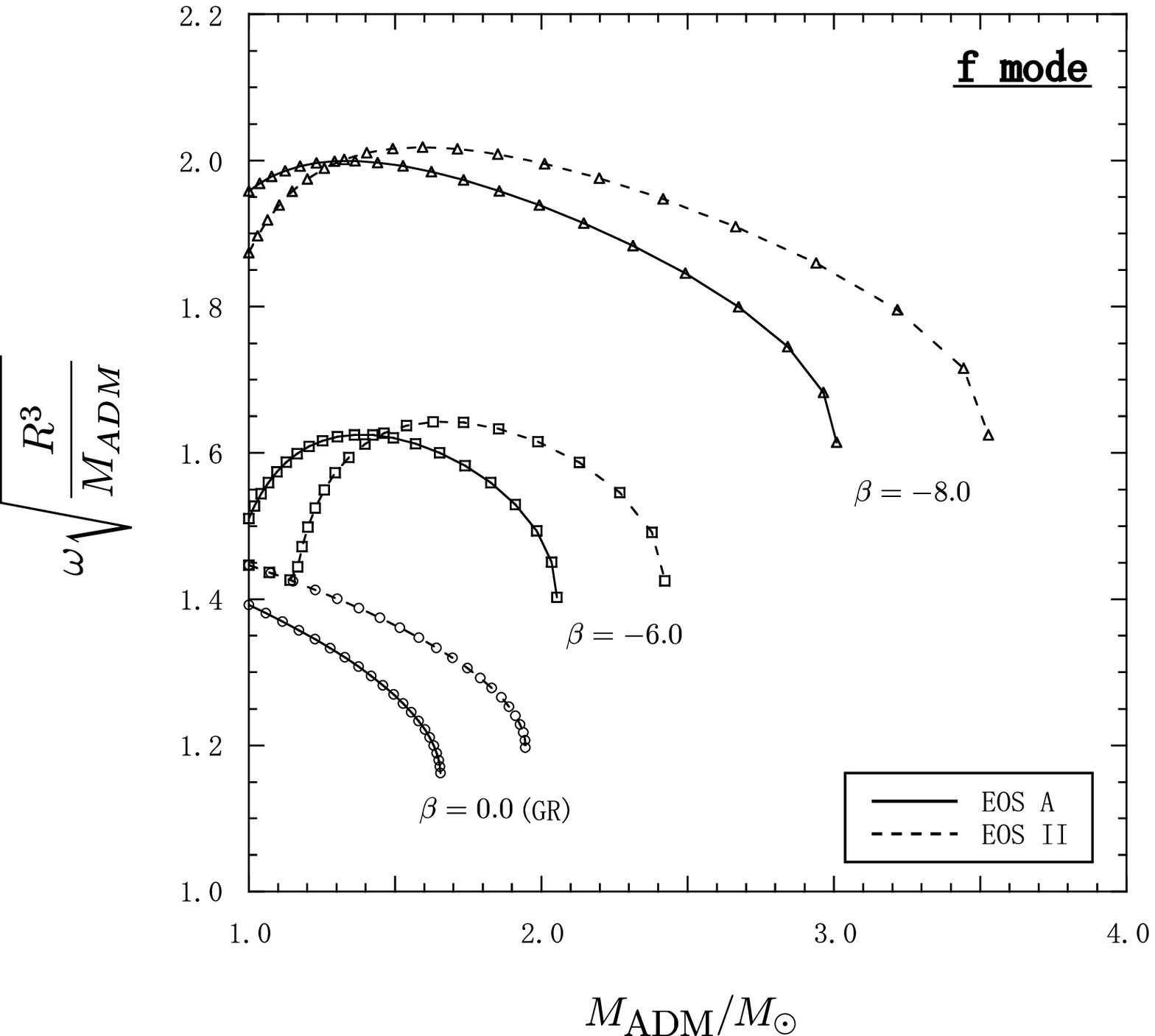}
\includegraphics[height=7cm,clip]{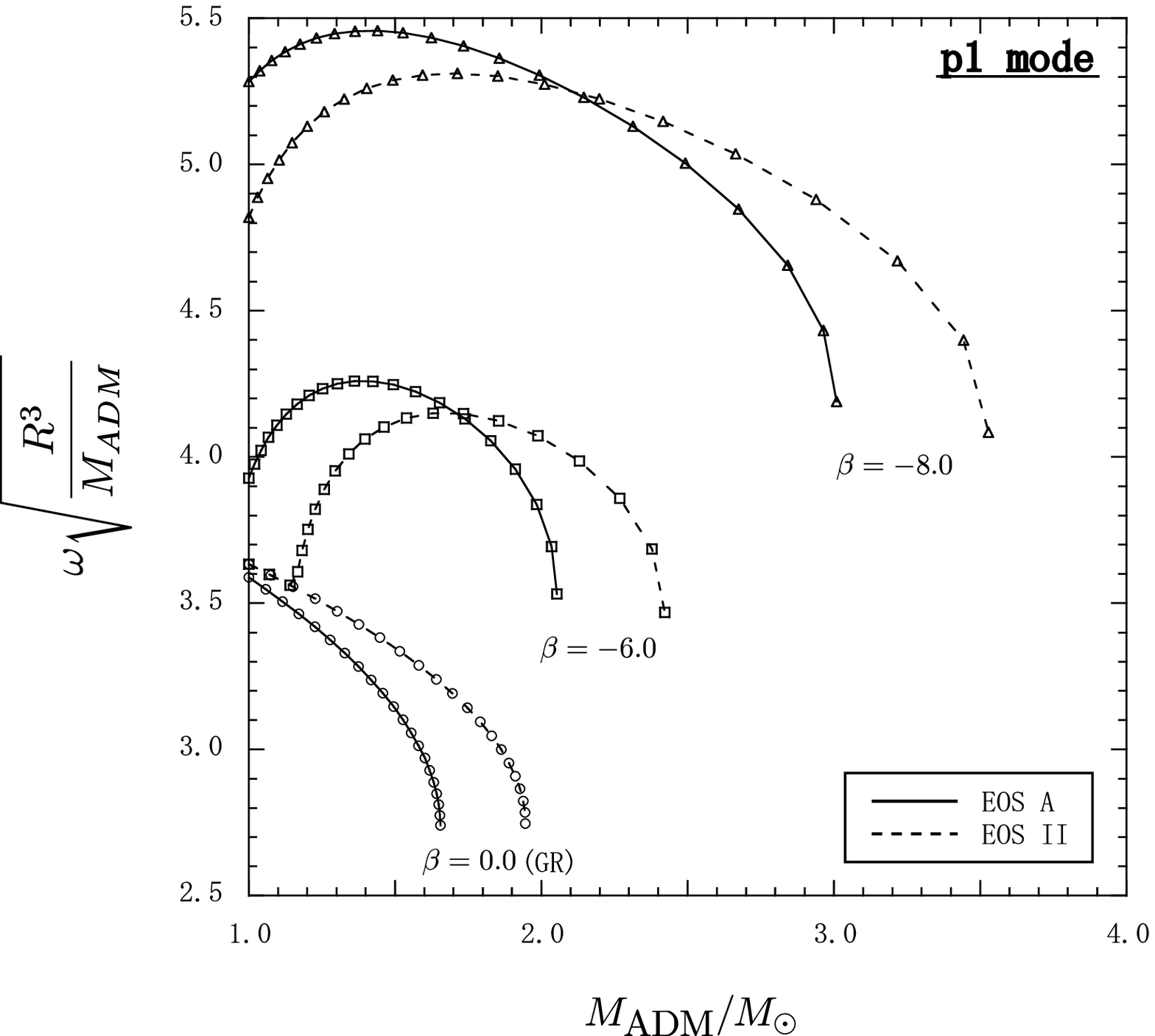}
\includegraphics[height=7cm,clip]{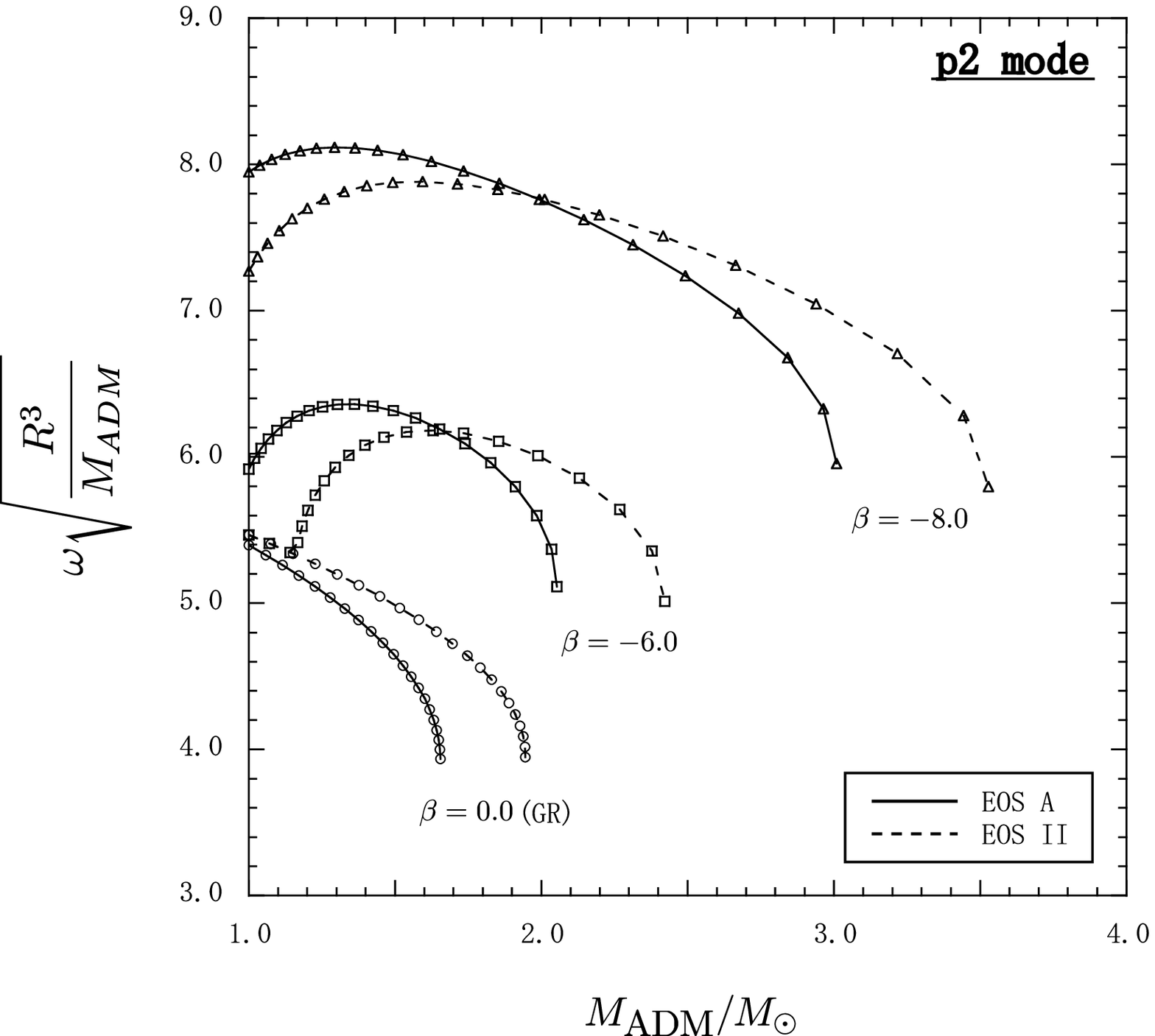}
\includegraphics[height=7cm,clip]{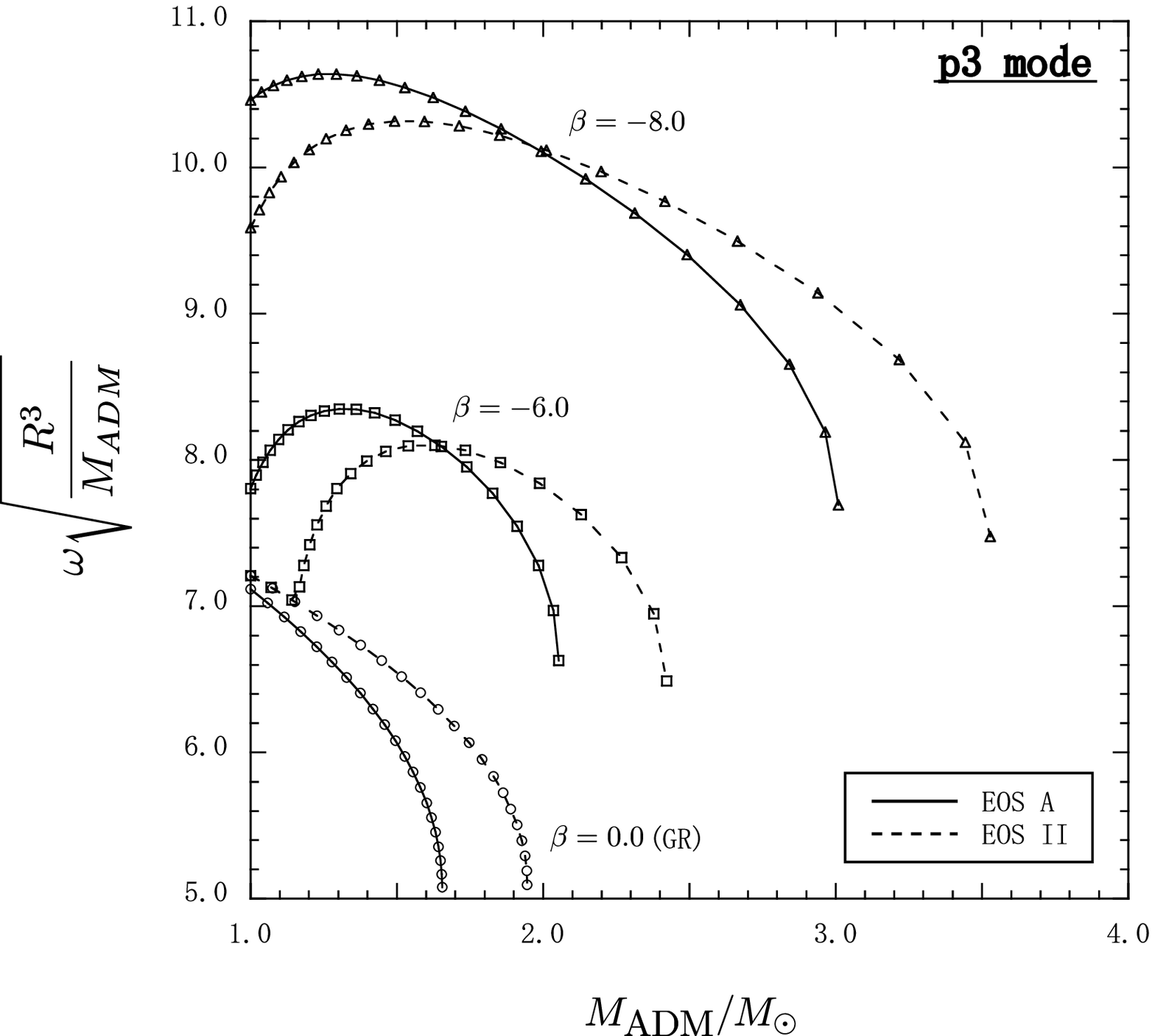}
\caption{The normalized frequencies of the first four fluid modes $f$ (upper-left panel), $p_1$ (upper-right panel, $p_2$ (lower-left panel), and $p_3$ (lower right panel) are plotted as functions of the ADM mass.  The results of the two EOS are shown for three values of the scalar parameter $\beta$ (0,-6 and -8). The asymptotic value of the scalar field is $\varphi_0=0$.  }
\label{fig:fig5}
\end{figure}

A careful study of Figure \ref{fig:fig4} suggests that the
observed higher frequencies predicted by scalar-tensor gravity
might be attributed to a denser neutron star. This possibility
cannot be ruled out if only one mode is observed, but as one can
easily find out from Figures \ref{fig_dependence-beta} and
\ref{fig:fig5} the effect of the scalar field on the higher fluid
modes, e.g. $p_1$, $p_2$ etc are more dramatic, and a synchronous
observation of a few modes will not only probe the presence of a scalar field but
it might provide a direct estimation of its strength. Concluding
the discussion related to Figure \ref{fig:fig4} we should admit
that there is no apparent explanation why the $f$-mode frequencies
after reaching a maximum either do not increase as the density
increases ($\beta=-6$) or even move towards lower frequencies
($\beta=-7.0$ and -8.0).

In the two preceding Figures  \ref{fig_dependence-beta} and
\ref{fig:fig4} we have shown the effect of the varying scalar
factor $\beta$ on the frequency of the modes. In Figure
\ref{fig_dependence-beta} the ADM mass and the asymptotic value of
$\varphi_0$ were fixed while in figure \ref{fig:fig4} we have
studied only one mode (the $f$-mode) for fixed $\varphi_0$. The
next figures show the effect of the scalar field on the
frequencies of the $f$, $p_1$, $p_2$, and $p_3$ modes for varying
ADM mass or/and the asymptotic value of $\varphi_0$. In Figure
\ref{fig:fig5} we plot the value of the normalized frequency of
the mode as function of the ADM mass.
One can easily observe that the differences due to the presence of the scalar field are
really impressive.
The models have been chosen to span a range of
masses from $1M_\odot$ up to the maximum allowed mass from each
theory. The normalized frequencies in these diagrams show
that the imprint of the scalar field is not only apparent but it can
influence the spectrum in a dramatic way. For example, for all
four oscillation modes ($f$, $p_1$, $p_2$, and $p_3$) the
normalized frequency of the maximum mass model in GR is {\em half}
of the corresponding frequency to an equal mass model with
$\beta=-8$ or about 40\%  smaller for $\beta=-7$. A very interesting observation
is that in contrast to the redshift of the atomic
lines\cite{DeDeo2003} which for the maximum mass models is
typically smaller than 10\% (see Tables I and II) the difference
in the frequencies between the models with $\beta=0$(GR) and
$\beta=-8$ is larger than 50\%. This figure suggests that a
possible observation of more than one mode can clearly signal the
existence of the scalar field.

\begin{figure}[htbp]
\centering
\includegraphics[height=8cm,clip]{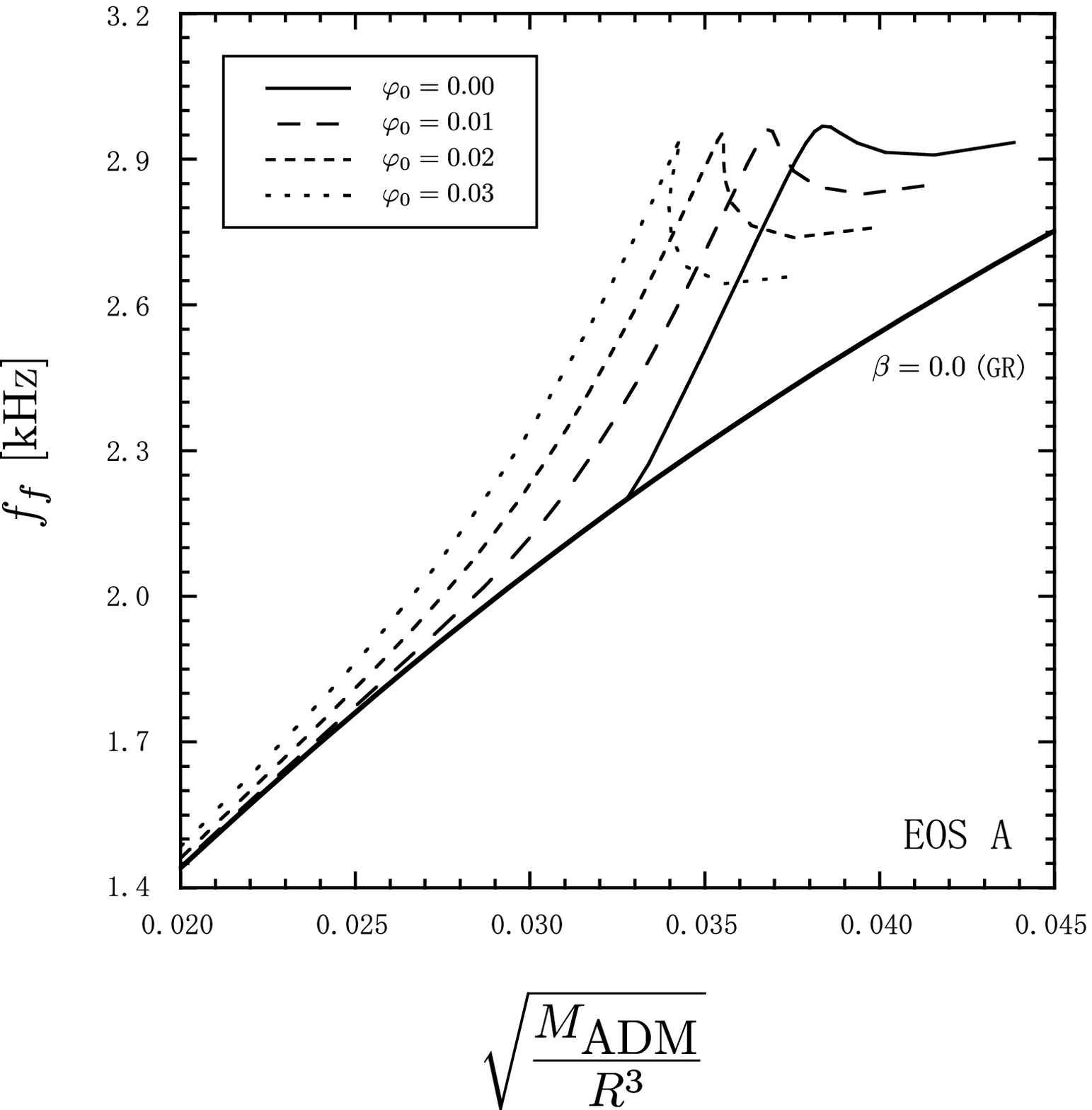}
\includegraphics[height=8cm,clip]{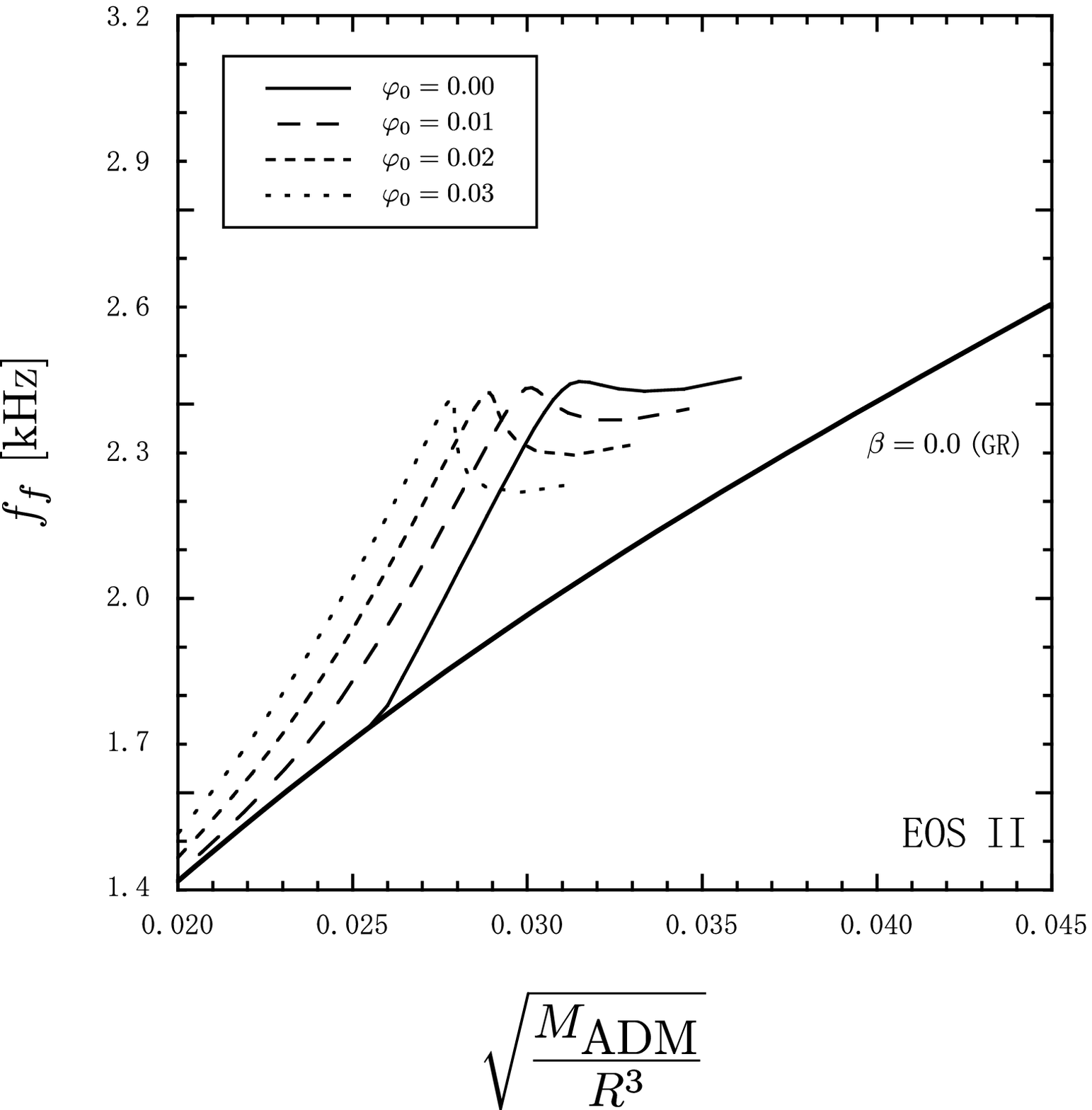}
\caption{The frequency of the $f$-mode as function of the averaged
density $(M_{\rm ADM}/R^3)^{1/2}$ of the star (notice that
$f_f=\omega_f/2\pi$). The thick solid line corresponds to  the
values of the mode for $\beta=0.0$ (GR), the thinner continuous and
dashed lines show the effect of the asymptotic value of the scalar
field $\varphi_0$ for a fixed value of the scalar parameter
$\beta=-6.0$. The left panel corresponds to EOS A and the right
panel to EOS II. } \label{fig:fig6}
\end{figure}

\begin{center}
\begin{figure}[htbp]
\includegraphics[height=8cm,clip]{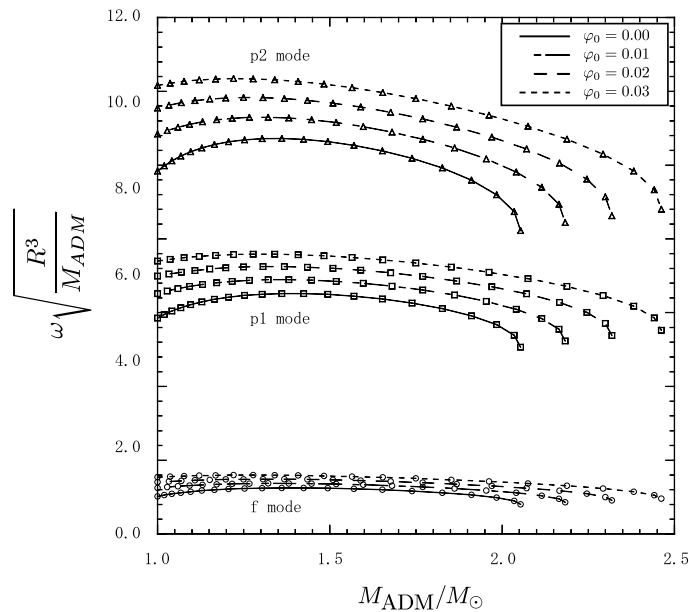}
\caption{
The dependence of eigenfrequencies on the value of $\varphi_0$,
where it is fixed $\beta=-6$ for EOS A. The range of ADM mass is
from $1.0M_{\odot}$ to maximum ADM mass for each $\varphi_0$. The
open circles, squares and triangles represent the $f$, $p_1$ and
$p_2$ modes, respectively. }
\label{fig_dependence-varphi_0}
\end{figure}
\end{center}

In all previous discussion about the effect of the scalar field on
the eigenmode frequencies of the star we have shown results for
stellar models for which the asymptotic value of the scalar field
was assumed zero. It is still an important question whether the
asymptotic value of $\varphi$ can be traced via the neutron star
asteroseismology. The results of a varying $\varphi_0$ are shown in
Figures \ref{fig:fig6}, \ref{fig_dependence-varphi_0}. Figure
\ref{fig:fig6}, is similar to Figure \ref{fig:fig4} but here we
have fixed the scalar factor to $\beta=-6$ and we vary
$\varphi_0$. In figure \ref{fig_dependence-varphi_0} we show the
dependence of $f$, $p_1$ and $p_2$ modes for varying $\varphi_0$
for the EOS A, the results for EOS II are similar. The last two
figures show that the imprint of an asymptotically non-vanishing
scalar field can be observed, while its becomes more pronounced for the higher modes.
Varying $\varphi_0$ from 0 to 0.03 we observe frequency variations larger
than 10-20\%, variations of this order can provide additional
constraints on the  asymptotic value of the scalar field.

%
%
%
%

\section{Conclusions}
\label{sec:conclusions}

In this paper we have  discussed the effect of the scalar-tensor theories on the oscillation spectra of neutron stars.
Scalar-tensor theories of gravity are generalizations of general
relativity and provide a natural connection to superstring
theories \cite{Callan1985,Damour1994b} as well as to inflation
\cite{La1989}. The presence of a scalar field in the
the strong-field regime has already been constrained by weak field
experiments and here we provide an additional way of testing/constraining their
existence via the gravitational wave asteroseismology.

We show that the oscillation frequencies of neutron stars carry
very ``clean'' imprints of the presence of the scalar field. An
observation of the neutron star oscillation spectrum via
gravitational waves or via electromagnetic signals emanating from
or around the surface of a neutron star will not only probe the
existence of the scalar field but it might also provide a
measurement of its asymptotic value.

In this study we have used the Cowling approximation which,
although  a restricted way of studying stellar oscillation, has
proven to be a very accurate and useful tool in asteroseismology.
Still, more detailed study is needed for proper modeling of the
effect. The inclusion of metric and scalar field perturbations
will result to additional information in the gravitational wave
spectrum. The combination of this extra information should provide
more accurate constraints on the existence of the scalar field.

\acknowledgments

We would like to thank G.Esposito-Far\'ese for his constructive 
comments and suggestions which improved our understanding of the problem. 
We would also like to thank N.Andersson, K.Maeda,
S.Mizuno and N. Stergioulas for useful discussions.
This work was partially supported by a Grant for The 21st Century
COE Program (Holistic Research and Education Center for Physics
Self-Organization Systems) at Waseda University. KK acknowledges
the hospitality of the Institute of Mathematics at the University
of Southampton. This work was also partially supported through the
Center of Gravitational Wave Physics, which is funded by the NSF
number cooperative agreement PHY 01-14375.



\begin{thebibliography}{999}


\bibitem{Fierz1956}
    M.Fierz
    Helv. Phys. Acta {\bf 29}, 128 (1956)

\bibitem{Jordan1959}
    P. Jordan
    Z.Phys. {\bf 157}, 112 (1959)

\bibitem{Brans1961}
    C. Brans and R. H. Dicke,
    Phys. Rev. {\bf 124}, 925 (1961).

\bibitem{Damour1992}
    T. Damour and G. Esposito-Far\`ese,
    Class. Quantum Grav. {\bf 9}, 2093 (1992).

\bibitem{Will1993}
    C. M. Will,
    {\em Theory and Experiment in Gravitational Physics}
    (Cambridge University Press, Cambridge, England 1993).

\bibitem{Will2001}
    C.M.Will
    Living Rev. Relativ. {\bf 4}:4 (2001),
    available at http://www.livingreviews.org/Articles/Volume4/2001-4.

\bibitem{Damour1994a}
        T.Damour, A.M.Polyakov
    { Nucl. Phys. }{\bf B 423}, 532 (1994)

\bibitem{Damour1994b}
        T.Damour, A.M.Polyakov
    { Gen. Relativ. Gravit. }{\bf 26}, 1171 (1994)

\bibitem{Esposito2004}
    G.Esposito-Farese
    gr-qc/0402007
          
\bibitem{Damour1993}
    T. Damour and G. Esposito-Far\`ese,
    Phys. Rev. Lett. {\bf 70}, 2220 (1993).

\bibitem{Bertotti2003}
       B. Bertotti, L. Iess, P. Tortora
       Nature {\bf 425} 374 (2003)    

\bibitem{Wiaux-1} J-M. G\'erard and Y. Wiaux, {Phys. Rev. D} {\bf 66}, 024020 (2002)

\bibitem{Wiaux-2} V. Boucher, J-M. G\'erard, P. Vandersheynst and Y. Wiaux astro-ph/0407208 
\bibitem{Damour1996}
    T. Damour and G. Esposito-Far\`ese,
    Phys. Rev. D {\bf 54}, 1474 (1996).

\bibitem{Harada1998}
    T. Harada,
    Phys. Rev. D {\bf 57}, 4802 (1998).

\bibitem{Damour1998}
    T. Damour and G. Esposito-Far\`ese,
    Phys. Rev. D {\bf 58}, 042001 (1998)

\bibitem{Scharre2002}
    P.D. Scharre and C.M. Will,
    Phys. Rev. {\bf D 65}, 042002 (2002)

\bibitem{Shibata1994}
    M.Shibata, K.Nakao, T.Nakamura
    Phys. Rev. D {\bf 50} 6058 (1994)

\bibitem{Scheel1995}
    M.A. Scheel, S.L. Shapiro, S.A. Teukolsky
    Phys. Rev. D {\bf 51}, 4236 (1995)

\bibitem{Harada1997}
    T.Harada, T.Chiba, K.Nakao, T.Nakamura
    Phys. Rev. D {\bf 55} 2024 (1997)

\bibitem{Saijo1997}
    M.Saijo, H.Shinkai, K.Maeda
    Phys. Rev. D {\bf 56} 785 (1997)

\bibitem{Novak1998}
    J. Novak
    Phys.Rev. D {\bf 57} 4789 (1998)

\bibitem{Novak2000}
        J. Novak, J. M. Ibanez
    Astroph. J. {\bf 533} 392 (2000)

\bibitem{Will2004}
	C.M. Will and N. Yunes
	gr-qc/0403100
    
\bibitem{DeDeo2003}
    S. DeDeo and D. Psaltis,
    Phys. Rev. Lett. {\bf 90}, 141101 (2003).

\bibitem{DeDeo2004}
    S. DeDeo and D. Psaltis,
    astro-ph/0405067.

\bibitem{Andersson1996}
        N. Andersson,  K.D. Kokkotas,
    Phys.~Rev.~Letters, {\bf 77}, 4134 (1996).

\bibitem{Andersson1998}
        N. Andersson,  K.D. Kokkotas,
    MNRAS, {\bf 299}, 1059 (1998).

\bibitem{Kokkotas2001}
        K. D. Kokkotas, T. A. Apostolatos and  N. Andersson,
    MNRAS, {\bf 320 }, 307 (2001)


\bibitem{Andersson2001a}
    N. Andersson and G. Comer,
    Phys. Rev. Let {\bf 87}, 241101 (2001)

\bibitem{Kokkotas2004}
    K.D. Kokkotas, P. Laguna, C. Sopuerta
    {\em preprint} (2004)

\bibitem{Sotani2003}
    H. Sotani and T. Harada,
    Phys. Rev. D, {\bf 68}, 024019 (2003).

\bibitem{Sotani2004}
    H. Sotani, K. Kohri and T. Harada,
    Phys. Rev. D, {\bf 69}, 084008 (2004).

\bibitem{Arnett1977}
    W. D. Arnett and R. L. Bowers,
    Astrophys. J. Suppl. {\bf 33}, 415 (1977).

\bibitem{Diaz1985}
    J. Diaz-Alonso and J. M. Iba\~nez-Cabanell,
    Astrophys. J. {\bf 291}, 308 (1985).

\bibitem{DN1993}
     T. Damour and K. Nordtvedt
     Phys. Rev. D, {\bf 48}, 3436 (1993)
     
\bibitem{Comer1998}
     G. L. Comer and H. Shinkai
     Class. Quantum Grav., {\bf 15}, 669 (1998)
              
\bibitem{Cottam2002}
    J.Cottam, F.Paerels, M.Mendez
    Nature {\bf 420}, 51 (2002)

\bibitem{Kokkotas1992}
    K.D.Kokkotas, B.F.Schutz
    MNRAS {\bf 255}, 119 (1992)

\bibitem{Callan1985}
C.G. Callan, E.J. Martinec, M.J. Perry and D. Friedan, Nucl.Phys.B
{\bf 262}, 593 (1985)


\bibitem{La1989}
    D.La and P.J.Steinhardt
    Phys. Rev. lett. {\bf 62}, 376 (1989)


\end{thebibliography}
\end{document}